# Gamma-ray production cross sections in proton interactions with $^{nat}$Mg, $^{nat}$Si and $^{56}$Fe targets: measurement over the energy range of $E_p$ = 66 - 125 MeV, data analysis, results and discussion. Astrophysical implications.

M. Debabi[1], Saad Ouichaoui[1*], A. Belhout[1], J. Kiener[2], E. A. Lawrie[3]. J. J. Lawrie[3,6], D. Moussa[1], W. Yahia-Cherif[3], Y. Rahma[4], H. Benhabiles-Mezhoud[5], A. Trabelsi[4], T. D. Bucher[3,11], A. Chafa[1], L. J. Conradie[3], S. Damache[4], I. Deloncle[2], J. L.Easton[6], C. Hamadache[2], F. Hammache[7], P. Jones[3], V. B. Kheswa[8], N. A. Khumalo[9], A. Kunyana[10], T. Lamula[9], S. N. T. Majola[11], J. Ndayishimye[8], D. Negi[3], S. P. Noncolela[9], S. Ouziane[1], P. Papka[3], Nicolas de Séréville[7], J. F. Sharpey-Schafer[6], O. Shirinda[8], V. Tatischeff[2], M. Wiedeking[3], and S. Wyngaardt[8]

[1]University of Sciences and Technology Houari Boumediene (USTHB), Faculty of Physics, P.O. Box 32, EL Alia, 16111 Bab Ezzouar, Algiers, Algeria
[2]Centre de Sciences Nucléaires et de Sciences de la Matière (CSNSM), CNRS-IN2P3 et Université de Paris-Sud, 91405 Orsay Campus, France
[3]iThemba LABS, National Research Foundation, P.O. Box 722, Somerset West 7129, South Africa
[4]CRNA, 02 Boulevard Frantz Fanon, B.P. 399 Alger-gare, Algiers, Algeria
[5]University M'Hamed Bougara, Institute of Electric and Electronics Engineering, 35 000 Boumerdès, Algeria
[6]Department of Physics, University of the Western Cape, Private Bag X17, Bellville 7535, South Africa
[7]Institut de Physique Nucléaire (IPN), CNRS-IN2P3 et Université Paris-Sud, 91405 Orsay Campus, France
[8]Department of Physics, Stellenbosch University, Private Bag X1, Matieland 7602, South Africa
[9]Department of Physics, University of Zululand
[10]Department of Physics and Engineering University of Zululand Private Bag X1001, Kwa-Dlangezwa 3886 South Africa
[11]Department of Physics, University of Cape Town, Private Bag X3, 7701 Rondebosch, South Africa

## Abstract:

We have measured nuclear γ-ray line production cross sections in interactions of highly accelerated proton beams with various target nuclei abundant in astrophysical sites. The experiments were carried out at the k = 200 Separated Sector Cyclotron (SSC) of iThemba LABS (near Cape Town, in South Africa) using a high-energy resolution and high efficiency detection system for registering the emitted γ-ray photons. We report and discuss in this paper the collected experimental data sets for various γ-ray lines produced in bombarding natural Mg ($^{nat}$Mg), natural Si ($^{nat}$Si) and $^{56}$Fe targets with proton beams of incident energies of $E_p$ = 66, 80, 95, 110 and 125 MeV. After describing the experimental set up and the data analysis method used, we report and discuss our results of experimental total cross sections in comparison to previous counterparts from the literature, to a semi-empirical compilation and to the predictions of nuclear reaction theory via performing comprehensive TALYS code calculations. Significantly improved agreements between theory and experiment are pointed out when using our modified optical model potential (OMP) and $\beta_\lambda$ level deformation parameters instead of the default input parameters built in TALYS.

Finally, we discuss and put into perspective the applications of our results in nuclear physics and related interdisciplinary fields, γ-ray astronomy and astrophysics with drawing relevant conclusions.

Keywords: Proton-induced nuclear reactions; γ-ray production cross sections; γ-ray spectrometry; γ-ray spectroscopy; Astrophysical implications

## 1. Introduction

Nuclear γ-ray production cross sections in interactions of highly accelerated particles (protons, deuterons, $^{3}He^{+}$ ions, α-particles) with atomic nuclei are needed for several purposes in various scientific research fields such as nuclear physics, proton therapy, materials science, γ-ray astronomy and astrophysics. In particular, measuring these observables at ground laboratory ion accelerators bears great importance to the latter two disciplines for simulating and modeling the γ-ray line emission in astrophysical sites like solar flares (SFs), the interstellar medium (ISM), supernovae remnants (SNRs), and cosmic compact objects (CCOs: pulsars, quasars, active galactic nuclei, etc.). The obtained results can allow the subsequent validation of observational data reported by space telescopes like the e-ASTROGAM [1,2], FERMI [3] and precursor space missions (SMM [4,5], CGRO [5], RHESSI [6] and INTEGRAL [7, 8]).

Under non-thermal acceleration by strong magnetic fields in astrophysical sites charged particles, like cosmic rays (CRs) and low-energy cosmic rays (LECRs), can reach very high energies reaching several hundreds of GeV/n. They violently interact with (H, He, C, N, O, Ne, Mg, Al, Si, S, Ca and Fe) abundant elements, giving rise to various nuclear reaction processes including the de-excitations of synthesized radioactive nuclei and excited nuclei upon undergoing (inelastic scattering, nucleon transfer, charge-exchange, pre-equilibrium and/or spallation) nuclear reactions. These reaction processes produce discrete γ-ray lines that can provide valuable information (nature, sources, spatial distribution, energy spectra and relative abundances) on the properties of the accelerated particles [9-13] like, e.g., the LECRs, as well as the composition of the accelerating media. Due to their deviation and spreading by the solar winds, the latter particles, that permanently bombard the inner galaxy, are elusive to direct observation. This is particularly the case of the protons and α-particles from SFs and the ISM whose energies extend from the nuclear reaction thresholds up to very high values out of reach in ground laboratory ion accelerators. The most important nuclear reactions are induced by the latter particles on heavier nuclei of abundant chemical elements and the inverse reactions between the accelerated ions of the these elements and the protons and α-particles from the dominant Hydrogen and Helium [14–18]. Three specific γ-ray line components have been identified [19] in the observed γ-ray line spectra from SFs and the ISM, consisting in a narrow component, a broad component of strong, discrete γ-ray lines, and a third non-resolved component of weak γ-ray lines appearing in the form of a quasi-continuum emitted by highly excited nuclear states.

Therefore, laboratory measurements of γ-ray line production cross sections over as wide as possible particle energy domains is required for elucidating the complex nuclear reaction processes at work in astrophysical sites, and for understanding the subsequent γ-ray line emission and associated complex energy spectra.

Previous measurements of these observables have been carried out mainly at the tandem accelerators of the Washington [20 - 23] and Orsay-Paris-Saclay [17, 24-28] universities for proton and α-particle beams with incident energies reaching values of $E_p$ = 26 and $E_\alpha$ = 40 MeV respectively. Besides, experimental cross section data for some strong γ-ray lines have been taken at the Lawrence Berkeley Laboratory's 88-Inch cyclotron accelerators [29] for these particles accelerated, respectively, up to $E_p$ = 50 MeV and $E_\alpha$ = 40 MeV. More recently, Kiener et al. [30] extended the experimental cross section data at the Berlin's Helmholtz-Zentrum cyclotron accelerator for the strongest γ-ray lines produced in α-particle induced reactions on C, Mg, Si and Fe targets over the incident energy range of $E_\alpha$ = 50 - 90 MeV. In another hand, a semi-empirical compilation of experimental γ -ray cross section data sets available in 2009 for strong lines prominent in SFs and the ISM exists [31]. It also predicts values for higher particle energies not covered by experiment, based only on TALYS code [32] calculations.

In this context, we have undertaken an experimental campaign in joint scientific collaboration (see, e.g., Ref. [18]) aiming at measuring γ-ray line production cross sections at the SSC facility of iThemba LABS. We have investigated a collection of elemental solid targets of natural C ($^{nat}$C), C and O (in form of Mylar), $^{nat}$Mg, $^{nat}$Si, $^{40}$Ca and $^{56}$Fe bombarded by high-energy proton beams. Given that earlier experimental data sets were available mainly for proton energies of $E_p \cong 26$ MeV [17, 20-28], we explored the remaining higher energy region of $E_p$ = 30 - 200 MeV permitted by the iThemba LABS's cyclotron accelerator. We successively carried out measurements at three separate proton energy ranges of $E_p$ = 30 - 66 MeV in 2014, $E_p$ = 66 - 125 MeV in 2015 and $E_p$ = 125 - 200 MeV in 2016 using the AFRODITE Germanium array for detecting the emitted γ-rays [33 - 35]. In performing this program, the measurements at the proton boundary energies of $E_p$ = 66 and 125 MeV were repeated for the sake of data checking and validation.

Part of the derived nuclear γ-ray line production cross sections has been the subject of several publications including those reported in Refs. [18, 36-39]. In particular, we extended earlier experimental data for main intense γ-ray lines from the $^{nat}Mg$, $^{nat}Si$ and $^{56}Fe$ targets to higher proton energies up to $E_p$ = 66 MeV and reported new experimental cross section data for weaker lines in Ref. [18], while in Refs. [36 - 38] we reported new experimental cross section data and shape analysis for de-excitation lines produced in proton-induced nuclear reactions on $^{nat}C$ and Mylar targets over the incident energy range of $E_p$ = 30 - 200 MeV. Finally, in Ref. [39] we optimized the OMP parameters for nucleon-induced reactions on $^{40}Ca$, calculated the γ-ray production cross sections for intense lines from the $^{36,38}Ar$ isotopes that we discussed in comparison to our experimental data measured over the incident proton energy range of $E_p$ = 30 – 125 MeV. In the present paper, we report on new experimental cross section data for γ-ray lines produced in interactions of 66, 80, 95, 110 and 125 MeV proton beams with the $^{nat}Mg$, $^{nat}Si$ and $^{56}Fe$ solid targets.

In the remaining of this report, we first present the experimental set-up and related methods used in section 2, then we expose our procedure of data analysis in section 3. The derived experimental total, angle-integrated γ-ray line production cross section results are reported in Section 4 where they are discussed in comparison to counterpart data sets from the literature [17, 18, 24, 25, 29] and to the Murphy et al. semi-empirical compilation [31]. In section 5, we discuss our experimental total cross section results within the predictions of nuclear reaction theory via comprehensive TALYS code [32] calculations. Finally, we discuss in section 6 the potential applications of our results, notably their astrophysical implications with drawing conclusions and emphasizing future work perspectives.

## 2. Experimental set-up and methods

Detailed descriptions of the experimental set-up and methods used in our experiments have been provided in our previous papers [18, 36]. We therefore limit ourselves here to presenting only the main related information.

Proton beams of energies $E_p$ = 66, 80, 95, 110, 125 MeV and current intensities of I = 2 - 5 nA delivered by the SSC were directed onto self-supporting solid targets of $^{nat}Si$, $^{nat}Mg$ and $^{56}Fe$ of thicknesses lying within the range of 5 - 9 mg/cm$^2$. Table 1 lists the explored targets and reports their characteristics and the corresponding bombarding proton beam energies.

Table 1: target properties and proton beam energies correspondingly used.

| Target | Thickness (mg/cm$^2$) | Purity (%) | Proton energies Ep (MeV) | References |
|---|---|---|---|---|
| Mg | 5 ± 0.02 | Natural[a] | 66, 80, 95 and 110 | [11-14] |
| | 8.99 ± 0.04 | Natural | 125 | |
| Si | 6 ± 0.04 | Natural[b] | 80, 95 and 110 | [11,12, 15, 16] |
| | 5.00 ± 0.03 | Natural | 66 and 125 | |
| Fe | 7.40 ± 0.07 | $^{56}$Fe(>99%) | 66, 80, 95, 110, 125 | [11-14] |
| | 5.50 ± 0.08 | $^{56}$Fe(>99%) | 125 | |

[a]Isotopic composition of $^{nat}$Mg = $^{24}$Mg (78.99%) + $^{25}$Mg (10%) + $^{26}$Mg (11.01%)

[b]Isotopic composition of $^{nat}$Si = $^{28}$Si (92.22%)+ $^{29}$Si (4.68%)+ $^{30}$Si (3.09%)

The purity of the targets in before the proton irradiations was higher than 99.99%. The targets were mounted onto rectangular Aluminum (Al) frames that were then fixed on a common Al ladder, equipped with a fluorescent $Al_2O_3$ viewer serving for tuning the proton beam and an empty-target frame devoted to measure the beam-induced γ-ray and neutron backgrounds.

The integrated beam current was measured to within 1% uncertainty by means of a well-designed, well-protected and electrically isolated Faraday cup located 3 meters away behind a concrete wall. The collected beam charge amounted typically to about 5 μC for 1-hour duration runs. The AFRODITE Ge array, composed of 8 clovers of the EUROGAM-phase II type [40] was used in fixed geometry for detecting the emitted γ-ray photons. Each clover contained 4 n-type HP-Ge crystals, each having dimensions of 50 x 70 mm$^2$. The HP-Ge crystals, shielded by means of Compton-suppressing BGO crystals, were housed into a common cryostat. The whole detection system subtended a large solid angle amounting to ~7.5% of the total 4π steradian sphere. 4 clovers were placed at laboratory angles of $\theta = 90^\circ$, while the other 4 were set at $\theta_{lab} = 135^\circ$ relative to the incident beam direction (see, e.g., Fig. 1 in Ref. [36]). The HP-Ge crystals in each clover were placed symmetrically at $\pm 5^\circ$ relative to its center. This configuration of the detection array, thus involving 32 HP-Ge crystals, allowed us to measure γ-ray angular distributions with only 4 observation angles each, i.e., $\theta_{lab} = 85^\circ, 95^\circ, 130^\circ$ and $140^\circ$. Besides, it was possible to place clover detectors forward at $\theta_{lab} = 45^\circ$ relative to the incident beam direction. However, in order to protect the HP-Ge crystals against probable damages due to possible high neutron fluxes from the proton-induced nuclear reactions on the targets and the Al and Ge elements composing the surrounding materials (beam pipe, reaction chamber, HP-Ge detectors), we preferred to avoid using these detection positions.

Energy calibration of the AFRODITE Ge array was done using standard radioactive sources of $^{60}$Co, $^{137}$Cs and $^{152}$Eu for covering the γ-ray low-energy range of $E_\gamma$ = 0.122 – 1.408 MeV. We also used the 6.129 MeV strong γ-ray line of $^{16}$O and associated escape lines at $E_\gamma$ = 5.107 and 5.618 MeV produced in the $^{16}$O (p, p'γ) inelastic proton scattering from the Mylar target. The relative energy resolution, $\Delta E_\gamma/E_\gamma$, frequently determined relative to the line of $^{60}$Co at $E_\gamma$ = 1332 keV, generally amounted to less than 0.20 %.

The same preceding radioactive sources were used for measuring the absolute detection efficiency, $\varepsilon\left(E_\gamma, \theta\right)$, versus the photon energy up to $E_\gamma$ = 1.408 MeV and the observation angle. Its values at fixed angle, $\theta$, were determined by the relation

$$\varepsilon\left(E_\gamma\right) = \frac{N(E_\gamma)}{A(t) \times t \times I} \qquad (1)$$

from the $N(E_\gamma)$ count rates (photopeak areas) in the γ-ray energy spectra, the $A(t)$ source activities at the measurement time, the acquisition time interval, $t$, and the γ-ray branching ratios, $I$, and were regularly checked during all the experimental runs. We extended this γ-ray energy range up to $E_\gamma$ = 3.5 MeV using efficiency values from a previous experiment [41] performed with a $^{56}$Co radioactive source in similar conditions to our measurements that we normalized to our experimental data. The systematic uncertainty in this normalization amounted to less than 2% based on the deviation in the region of overlap between the experimental data sets. Furthermore, we extrapolated the $\varepsilon\left(E_\gamma, \theta\right)$ values, as we did in all our previous experiments, to higher photon energies up to $E_\gamma \approx$ 10 MeV via performing Monte Carlo simulations of the AFRODITE detection array and surrounding materials (see next section).

The Gamma-ray background stems from various sources (the explored targets, the Al empty-target frame, the surrounding materials and the experimental room). We regularly recorded γ-ray background energy spectra: (i) without beam and (ii) with beam on an empty-target frame. Then, we used them to correct the count rates within the γ-ray energy spectra from the nuclear reactions of interest in the current study. The γ-ray count rates did not exceed 5-6 kHz per HP-Ge crystal, and were lower than 0.5 kHz during the runs performed with the beam passing through the empty-target frame. The beam current intensities did not exceed 5-6 nA in practically all the experimental runs, while the measured counting dead times were lying within the range of 6-8%.

A multi-instance data acquisition system (MIDAS [42]), composed of digital electronics modules made of XIA type "DGF Pixie-16", was used for collecting and storing the measured raw γ-ray event data. The experimental conditions were maintained stable with regularly tuning the proton beam, frequently monitoring the data acquisition system and checking the levels of the γ-ray and neutron backgrounds (see Refs. [18, 36] for details). Finally, we saved the measured γ-ray energy spectra and the related time information for the sake of subsequent off-line data analysis after the experiments.

## 3. Data analysis, γ-ray efficiency results, determination of γ-ray cross sections

The different steps in the treatment and analysis of the recorded γ-ray experimental data (energy spectra, their precise energy calibration, detection efficiencies) aim at determining as accurately as possible the γ-ray differential cross sections and the total, angle-integrated production cross sections. The analysis of the observed γ-ray line peaks of interest was carried out following the method detailed in Refs. [18, 24, 25, 36] applied to the 32 HP-Ge detectors used in our experiments. During the phase of data analysis following the experiments, we corrected the count rates within the raw γ-ray energy spectra for background events and dead time losses as detailed in Refs. [18, 36]. We subtracted the beam-induced background by removing the count rates measured with the proton beam passing through the empty-target frame after proper normalization to the same beam charge. The data analysis can be proceeded via two distinct modes: (i) the direct, *single mode* considering each HP-Ge crystal as an individual detector and (ii) the *add-back mode* consisting in summing the energies of coincident events in two adjacent HP-Ge crystals of the same clover. Given that the counting rate statistics in the γ-ray energy spectra was high enough, we have used the former mode for extracting the contents of the observed γ-ray line peaks of interest. Normally, the Compton scattering background is suppressed by means of the BGO crystals. However, for improving the suppression of this background component we additionally rejected all the γ-ray events observed in coincidence by two or more HP-Ge crystals of each clover detector. We used the ROOT [43] program and/or the GF3 software available in the RADWARE package [44] for extracting the line peak count rates in the useful γ-ray energy spectra. In general, when the γ-ray peaks were well defined and isolated as in the case of the energy spectra from the $^{56}$Fe target, we simply fitted symmetric Gaussian-shape distributions for extracting their content data (peak areas) with subtracting the related background by means of linear functions. However, a more elaborated multi-peak analysis method was applied in case of more complex, overlapping or significantly Doppler-broadened γ-ray line peaks as in case of the spectra from the $^{nat}$Mg and $^{nat}$Si targets. We evaluated the total relative uncertainties in the γ-ray line peak areas in form of three complementary contributions: (i) the systematic uncertainties, estimated to amount to 1 - 2% for strong line peaks and to 8 - 10% for least intense line peaks, (ii) the statistical uncertainties amounting to less than 2%, and (iii) the uncertainties in the fitting procedure not exceeding 3%.

The determination of the experimental γ-ray production cross sections requires the precise knowledge of the absolute detection efficiencies over an extended γ-ray energy interval. Therefore, as in our previous experiments [18, 36]), we further extrapolated the measured values of $\varepsilon\left(E_{\gamma},\theta\right)$ by means of the radioactive sources up to $E_\gamma \sim 10$ MeV via performing detailed Monte-Carlo simulations of the AFRODITE reaction chamber, the associated clover detection array and the surrounding materials using the GEANT4 software [45]. These simulations were made at the radioactive source energies and for regular γ-ray energy intervals between $E_\gamma = 80$ keV and $E_\gamma = 10$ MeV. Then, we normalized the simulated $\varepsilon\left(E_{\gamma},\theta\right)$ values to the experimental data taken by means of the calibrated radioactive sources over the lower photon energy range. Furthermore, we fitted the following multi-parametric function

$$\epsilon\left(E_{\gamma},\theta\right) = e^{[(A+BX+CX^2)^{-G}+(D+EY+FY^2)^{-G}]^{-1/G}}, \qquad (2)$$

reported in the RADWARE software package [44] to the derived efficiency data (see also Refs. [18, 36]). In this expression, $X = \log\left(E_\gamma/100\right)$, $Y = \log\left(E_\gamma/1000\right)$, and the A, B, C, D, E, F coefficients are free adjustable parameters. Fig. 1 reports, as a typical example, the variation of the absolute detection efficiency versus the photon energy for a single HP-Ge crystal set at $\theta_{lab} = 130°$. It shows the experimental efficiency data taken by the standard radioactive sources, the GEANT4-simulated values normalized to the latter data set and the fitted curve to both two data sets generated by the above multiparametric function. The fit curves to the efficiency data were finally used for calculating the efficiencies for all the 32 HP-Ge detectors and the explored targets, and were used for calculating the γ-ray line production cross sections. Other results for the absolute detection efficiencies have been reported and discussed in details in our previous papers with accounting for their proton energy dependence (see Refs. [18, 36]).

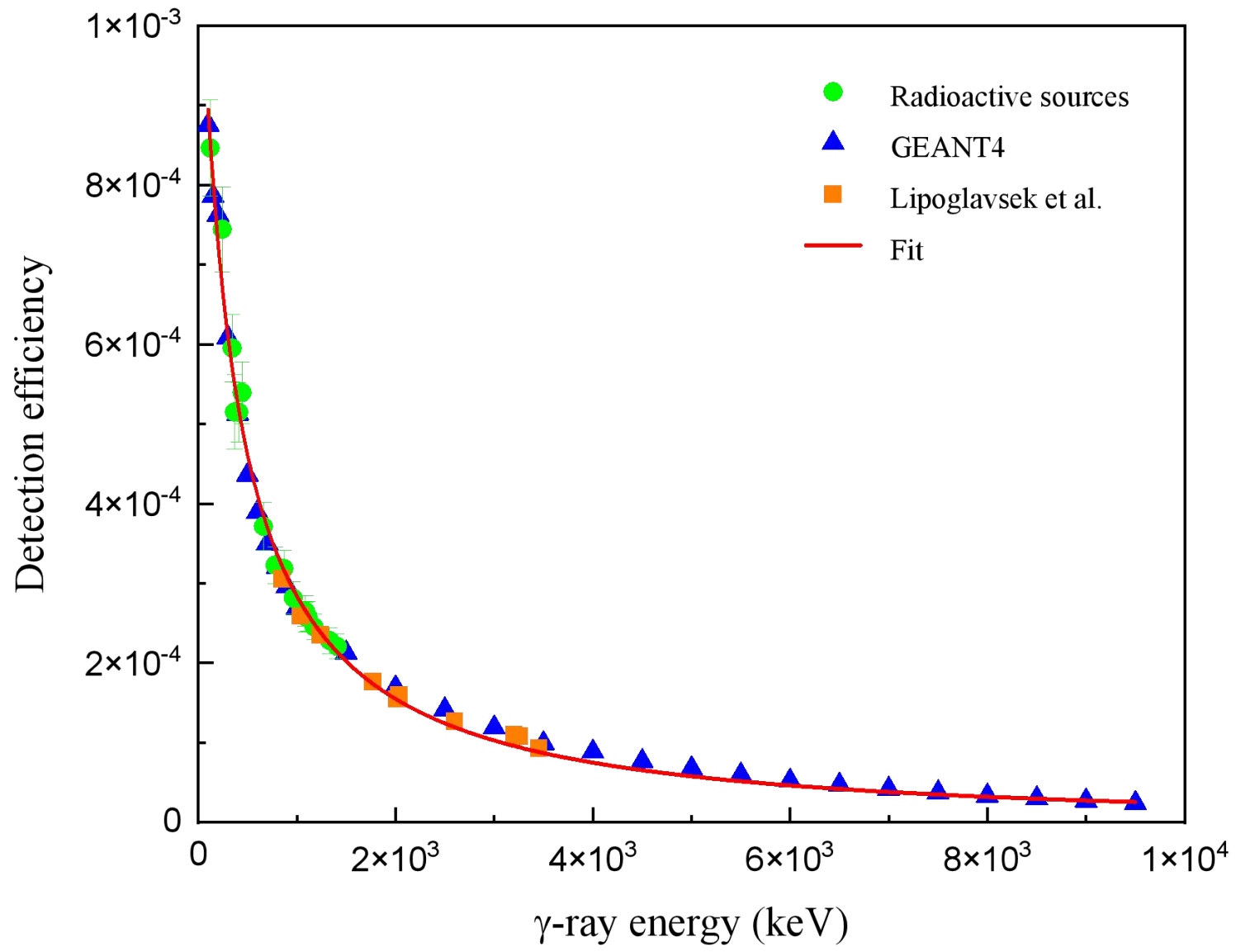

Fig. 1: absolute γ-ray detection efficiencies for a single HP-Ge crystal located at $\theta_{lab} = 130°$. The green circles represent the absolute efficiency data measured with the calibrated radioactive sources of $^{152}Eu$, $^{60}Co$ and $^{137}Cs$, the orange squares show the results obtained by Lipoglavšek et al [41], while the blue triangles correspond to the normalized GEANT4-simulated values. The fitted curve to these three data sets is shown by the red solid line.

In order to evaluate the uncertainties associated with the efficiency determination, we primarily considered the statistical uncertainty arising from the calculation of the areas under the full-energy absorption peaks. This statistical uncertainty was added in quadrature to the uncertainty in the activity of the radioactive sources and the uncertainty associated with the branching ratios. The uncertainty in the detection efficiency data taken by the radioactive sources was dominated by that affecting the radioactive source activities amounting to 4 – 5%, the branching ratios contributed by less than 1%, while the statistical uncertainty in the γ-ray line peak areas amounted to 1-2%. Thus, an average value of the overall uncertainty in $\varepsilon\left(E_{\gamma},\theta\right)$ from the radioactive sources of approximately 7% was obtained. Upon adding in quadrature the uncertainties associated with the GEANT4-simulated efficiencies, the normalization factors and the fits of the multi-parametric function to the data, the total uncertainty in $\varepsilon\left(E_{\gamma},\theta\right)$ increased up to ~12%.

The experimental γ-ray differential cross sections (angular distributions) were calculated using the following equation

$$\frac{d\sigma(E_\gamma,\theta)}{d\Omega} = \frac{N\,(E_\gamma,\ \theta)}{\epsilon(E_\gamma,\theta)N_t N_p}10^{27}\ (mb/sr), \qquad (3)$$

in terms of the γ-ray count rates, $N(E_\gamma, \theta)$, the absolute detection efficiencies, $\varepsilon$ ($E_\gamma$, $\theta$), the target thicknesses, $N_t$ (numbers of nuclei per cm$^2$) and the registered proton beam charge (number of projectiles hitting the targets), $N_p$. Then, we fitted a Legendre polynomial expansion of the form

$$W(\theta) = \sum_{\ell=0}^{\ell=max} a_\ell \, Q_\ell \, P_\ell \left(cos(\theta)\right) \quad (4)$$

to the differential cross section data. In this expression, the summation extends only over even integer $\ell$-values, $\ell_{max} = 2\lambda$ or 2J with λ and J being respectively the γ-ray multipolarity and the spin of the emitting nuclear state, and the $Q_\ell$ (ε, $\ell$, θ) are energy-dependent geometrical attenuation coefficients [46, 47] introduced for taking into account the finite dimensions of the γ-ray detectors and better evaluating the angular distribution expansion. We determined the latter coefficients via GEANT4 [45] Monte-Carlo simulations of the AFRODITE Ge detection array, finding nearly constant values of around unity over the whole photon energy range of 0.1 MeV ≤ Eγ ≤ 10 MeV of interest with relative uncertainties of ~1%. Except for two M3 γ-ray lines at 425.81 and 472.20 keV originating from isomeric states in $^{24}$Al and $^{24}$Na, respectively, the observed γ-ray transitions had λ-values of at most 2 (see next subsection), which led us to fix $\ell_{max} = 4$ in Eq. (4). For the former two transitions in $^{24}$Al and $^{24}$Na, we restricted the expansion to the zeroth order term, given that the long lifetimes of the isomeric states lead to nuclear de-orientation of the aligned states resulting in isotropic angular distributions [18].

Finally, we determined the total, angle-integrated γ-ray line production cross sections only in terms of the $a_0$ coefficients of the polynomial expansion given by Eq. (4), using the relation $\sigma = 4\pi a_0$.

## 4. Experimental results and discussion

### 4.1 γ-ray energy spectra, nuclear transitions

We performed precise identifications of the γ-ray line peaks with total absolute uncertainties amounting typically to less than 0.1 keV. The recorded γ-ray energy spectra at different proton beam energies showed great similarities in their shapes (γ-ray lines + backgrounds) and the presence of an intense γ-ray background line at $E_p = 511$ keV from the ($e^+$ - $e^-$) pair annihilation, and marked differences in their γ-ray line peak contents due to the differences in the target compositions. Fig. 2 reports typical examples of experimental γ-ray energy spectra recorded in the single mode by an HP-Ge detector positioned at $\theta_{lab} = 130^\circ$ for a 66 MeV proton beam interacting with the $^{nat}$Si target. Fig. 3 reports a net γ-ray energy spectrum from the same HP-Ge crystal corrected for the proton beam-induced background and the background from the reaction chamber.

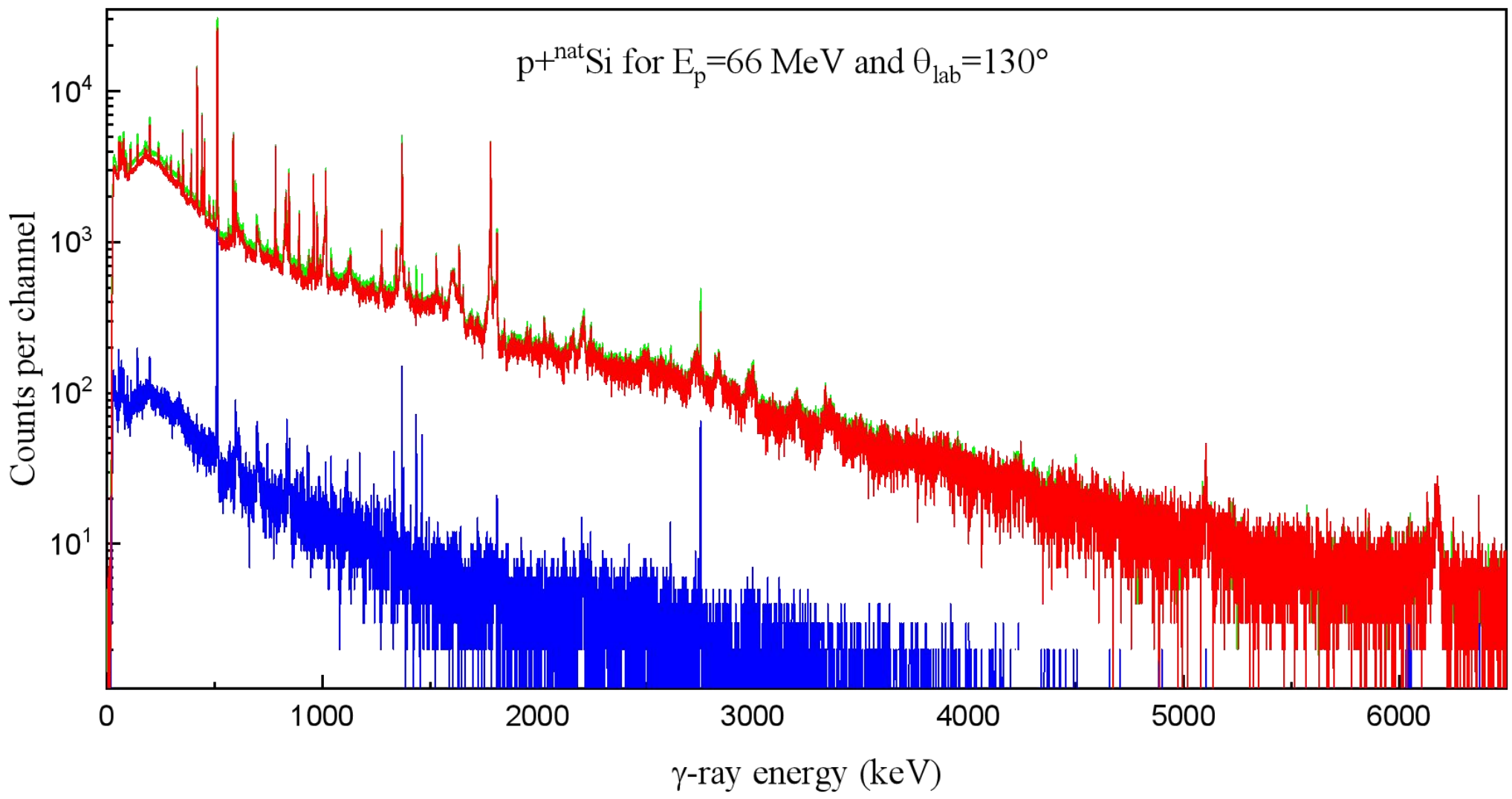

Fig. 2: γ-ray energy spectra recorded by an HP-Ge crystal positioned at $\theta_{lab}$= 130° in interaction of a 66 MeV proton beam with the $^{nat}$Si target. The total raw spectrum including the background with the target exposed to the proton beam is shown in green color. The background spectrum obtained with the proton beam passing through the empty-target frame (without target in place), normalized to the same accumulated beam charge, is depicted in blue color, while the net spectrum obtained after background subtraction is depicted in red.

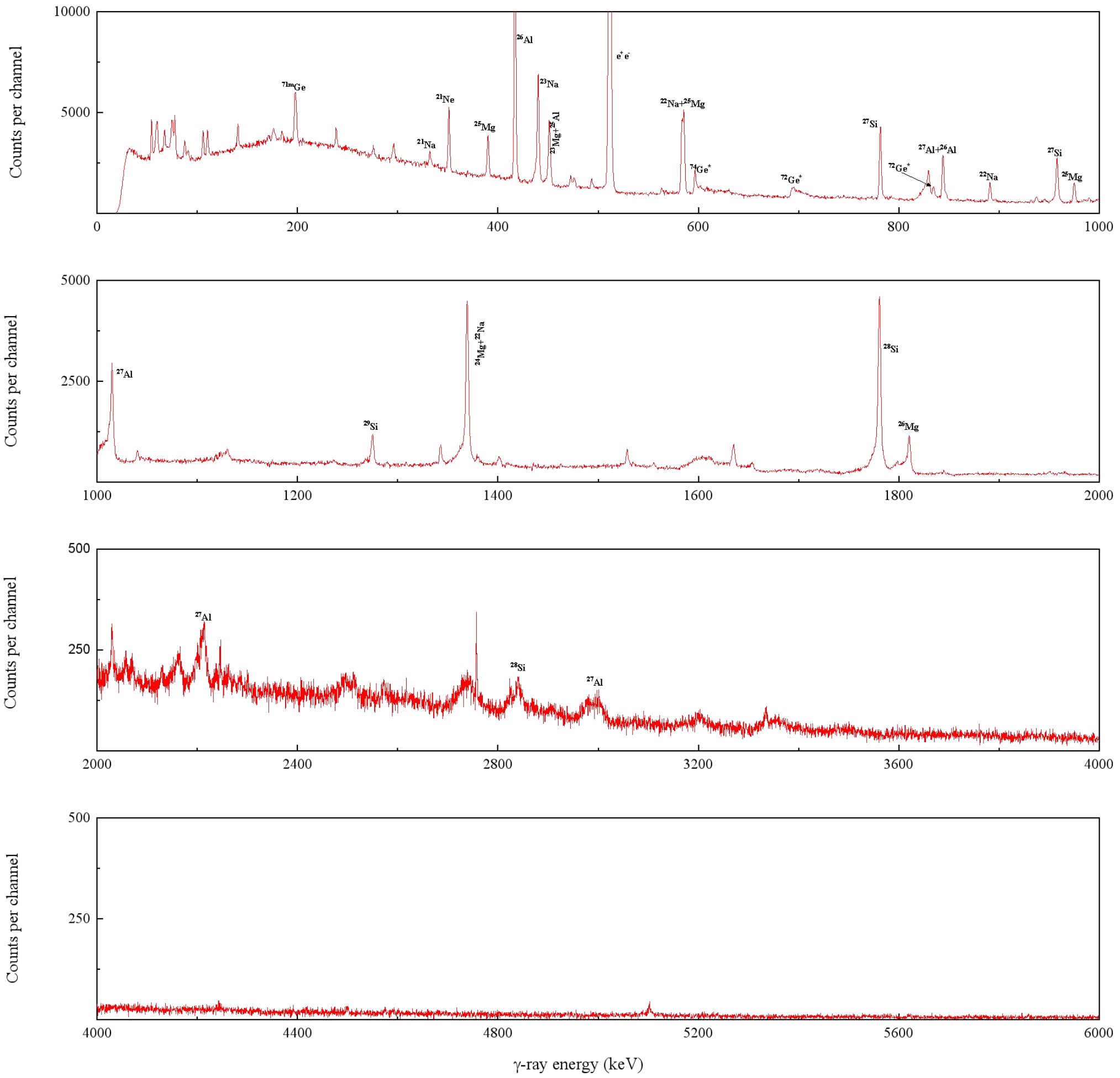

Fig. 3: net γ-ray energy spectrum from the same HP-Ge detector placed at $\theta_{lab}$= 130° in interaction of a 66 MeV proton beam with the $^{nat}$Si target obtained after subtracting the proton beam-induced background and the reaction chamber background, and further corrected for the Compton scattering background.

Numerous γ-ray lines of variable intensities are observed in Figs (2, 3). They result from the de-excitation of nuclear excited states produced in interactions of the incident proton beam with the target's constituent nuclei and the nuclei of the surrounding materials. Besides the intense 511 keV line, other background line peaks are present in the γ-ray energy spectra, stemming from proton and secondary neutron-induced reactions on the surrounding Al and Ge elements constituting the reaction chamber, the target frames and holder and the HP-Ge crystals [48, 49]. They include the line transition at $E_\gamma$ = 198 keV arising from the $^{70}$Ge (n, γ)$^{71m}$Ge radiative neutron capture reaction, the line at $E_\gamma$ = 595.85 keV produced in the $^{74}$Ge (n, n'γ)$^{74}$Ge$^*$ inelastic neutron scattering, and the lines at $E_\gamma$ = 689.60 and 834.01 keV originating from the $^{72}$Ge (n, n'γ)$^{72}$Ge$^*$ inelastic neutron scattering. The γ-ray line at $E_\gamma$ = 846.76 keV from the $^{56}$Fe (p, p'γ) $^{56}$Fe inelastic proton scattering is also observed, together with the 843.76 keV line of $^{27}$Al produced both via the $^{27}$Al (p, p'γ)$^{27}$Al$^*$ inelastic proton scattering and the β-decay of $^{27}$Mg following the $^{27}$Al (n, p) $^{27}$Mg$^{*}$ charge-exchange reaction. The neutron-related γ-ray background line peaks were found to affect the measured γ-ray energy spectra from all the targets explored in our experiments [18, 36].

The recorded γ-ray energy spectra from the $^{nat}$Mg, $^{nat}$Si and $^{56}$Fe targets generally exhibited rather narrow and well separated lines of various intensities less affected by Doppler effect (see Fig. 3), unlike the energy spectra from the Mylar and $^{nat}$C targets that contain strongly Doppler-shifted and broadened lines making their analysis more complex [36].

We list separately in Tables (2, 3, 4) the properties of the γ-ray line transitions identified in proton beam interactions with the explored $^{nat}$Mg, $^{nat}$Si and $^{56}$Fe targets, respectively. They correspond to the de-excitation of various isotopes produced mainly in (p, p') inelastic proton scattering and other nuclear reactions (pre-equilibrium, fusion-evaporation, etc.). As can be seen in these tables, the pointed out (Eλ, Mλ) nuclear γ-ray line transitions have essentially M1, E2 and (M1 + E2) characters with maximum multipolarity λ = 2.

Table 2. Properties of the γ-ray transitions observed in proton-induced nuclear reactions on the $^{nat}$Mg target.

| $E_\gamma$ (keV) | Intruder (keV) | Emitting nucleus | $E_i \rightarrow E_f$ (keV) | $J_i^\pi \rightarrow J_f^\pi$ | Multipolarity | Branching ratio (%) |
|---|---|---|---|---|---|---|
| 331.91 | | $^{21}$Na | 331.9 →g.s | $5/2^+ \rightarrow 3/2^+$ | M1+E2 | 100 |
| 350.73 | | $^{21}$Ne | 350.73 →g.s | $5/2^+ \rightarrow 3/2^+$ | M1+E2 | 100 |
| 389.71 | | $^{25}$Mg | 974.76 →585.05 | $3/2^+ \rightarrow 1/2^+$ | M1+E2 | 45.89 |
| 425.81 | | $^{24}$Al | 425.8 →g.s | $1^+ \rightarrow 4^+$ | M3 | 100 |
| 439.99 | | $^{23}$Na | 439.99 →g.s | $5/2^+ \rightarrow 3/2^+$ | M1+E2 | 100 |
| 450.7 | | $^{23}$Mg | 450.71 →g.s | $5/2^+ \rightarrow 3/2^+$ | M1+E2 | 100 |
| 472.2 | | $^{24}$Na | 472.21 →g.s | $1^+ \rightarrow 4^+$ | [M3] | 100 |
| 563.19 | | $^{24}$Na | 563.2 →g.s | $2^+ \rightarrow 4^+$ | [E2] | 3.61 |
| 585.03 | | $^{25}$Mg | 585.05 →g.s | $1/2^+ \rightarrow 5/2^+$ | E2 | 100 |
| | 583.04 | $^{22}$Na | 583.05 →g.s | $1^+ \rightarrow 3^+$ | E2 | 100 |
| 937.0 | | $^{18}$F | 937.2 → g.s | $3^+ \rightarrow 1^+$ | E2 | 100 |
| 974.74 | | $^{25}$Mg | 974.76 →g.s | $3/2^+ \rightarrow 5/2^+$ | M1+E2 | 54.11 |
| 1016.9 | | $^{23}$Ne | 1016.93 →g.s | $1/2^+ \rightarrow 5/2^+$ | M1+E2 | 100 |
| | 1016.8 | $^{22}$Na | 2968.6 →1951.8 | $3^+ \rightarrow 2^+$ | M1 | 100 |
| 1368.63 | | $^{24}$Mg | 1368.67 →g.s | $2^+ \rightarrow 0^+$ | E2 | 100 |
| | 1368.7 | $^{22}$Na | 1951.8 →583.05 | $2^+ \rightarrow 1^+$ | | 99.71 |
| 1808.68 | | $^{26}$Mg | 1808.74 →g.s | $2^+ \rightarrow 0^+$ | E2 | 100 |
| 2754.02 | | $^{24}$Mg | 4122.85 →1368.67 | $4^+ \rightarrow 2^+$ | E2 | 100 |

Table 3. Same as in Table 2 but for the γ-ray transitions observed in proton-induced nuclear reactions on the $^{nat}Si$ target.

| $E_\gamma$ (keV) | Intruder (keV) | Emitting nucleus | $E_i \rightarrow E_f$ (keV) | $J_i^\pi \rightarrow J_f^\pi$ | Multipolarity | Branching ratio (%) |
|---|---|---|---|---|---|---|
| 331.91 | | $^{21}Na$ | 331.9 →g.s | $5/2^+ \rightarrow 3/2^+$ | M1+E2 | 100 |
| 350.73 | | $^{21}Ne$ | 350.73 →g.s | $5/2^+ \rightarrow 3/2^+$ | M1+E2 | 100 |
| 389.71 | | $^{25}Mg$ | 974.76 →585.05 | $3/2^+ \rightarrow 1/2^+$ | M1+E2 | 45.89 |
| 416.85 | | $^{26}Al$ | 416.85 →g.s | $3^+ \rightarrow 5^+$ | [E2] | 100 |
| 439.99 | | $^{23}Na$ | 439.99 →g.s | $5/2^+ \rightarrow 3/2^+$ | M1+E2 | 100 |
| 450.7 | | $^{23}Mg$ | 450.71 →g.s | $5/2^+ \rightarrow 3/2^+$ | M1+E2 | 100 |
| | 451.7 | $^{25}Al$ | 451.7 →g.s | $1/2^+ \rightarrow 5/2^+$ | [E2] | 100 |
| 585.03 | | $^{25}Mg$ | 585.05 →g.s | $1/2^+ \rightarrow 5/2^+$ | E2 | 100 |
| | 583.04 | $^{22}Na$ | 583.05 →g.s | $1^+ \rightarrow 3^+$ | E2 | 100 |
| 780.8 | | $^{27}Si$ | 780.9 →g.s | $1/2^+ \rightarrow 5/2^+$ | E2 | 100 |
| 843.76 | | $^{27}Al$ | 843.76 →g.s | $1/2^+ \rightarrow 5/2^+$ | E2 | 100 |
| | 843.92 | $^{26}Al$ | 2913.4 →2069.47 | $2^+ \rightarrow (2^+)$ | [M1] | 67.76 |
| 890.87 | | $^{22}Na$ | 890.89 →g.s | $4^+ \rightarrow 3^+$ | M1+E2 | 100 |
| 957.3 | | $^{27}Si$ | 957.4 →g.s | $3/2^+ \rightarrow 5/2^+$ | (M1+E2) | 93.97 |
| 974.74 | | $^{25}Mg$ | 974.76 →g.s | $3/2^+ \rightarrow 5/2^+$ | M1+E2 | 54.11 |
| 1273.36 | | $^{29}Si$ | 1273.39 →g.s | $3/2^+ \rightarrow 1/2^+$ | M1+E2 | 100 |
| 1368.63 | | $^{24}Mg$ | 1368.67 →g.s | $2^+ \rightarrow 0^+$ | E2 | 100 |
| | 1368.7 | $^{22}Na$ | 1951.8 →583.05 | $2^+ \rightarrow 1^+$ | | 99.71 |
| 1778.97 | | $^{28}Si$ | 1779.03 →g.s | $2^+ \rightarrow 0^+$ | E2 | 100 |

Table 4. Same as in Table 2 but for the γ-ray transitions observed in proton-induced nuclear reactions on the $^{56}$Fe target.

| Eγ (keV) | Intruder (keV) | Emitting nucleus | $E_i \rightarrow E_f$ (keV) | $J_i^\pi \rightarrow J_f^\pi$ | Multipolarity | Branching ratio (%) |
|---|---|---|---|---|---|---|
| 125.95 | | $^{55}$Mn | 125.95 →g.s | $7/2^- \rightarrow 5/2^-$ | M1+(E2) | 100 |
| 156.27 | | $^{54}$Mn | 156.29 →g.s | $4^+ \rightarrow 3^+$ | M1+E2 | 100 |
| 211.98 | | $^{54}$Mn | 368.22 →156.29 | $5^+ \rightarrow 4^+$ | M1 | 99.1 |
| 226.3 | | $^{50}$V | 226.21 →g.s | $5^+ \rightarrow 6^+$ | M1+(E2) | 100 |
| 274.8 | | $^{55}$Fe | 2813.8 →2539.11 | $13/2^- \rightarrow 11/2^-$ | M1(+E2) | 100 |
| 377.88 | | $^{53}$Mn | 377.89 →g.s | $5/2^- \rightarrow 7/2^-$ | M1+E2 | 100 |
| 411.9 | | $^{55}$Fe | 411.42 →g.s | $1/2^- \rightarrow 3/2^-$ | M1+(E2) | 100 |
| | 411.4 | $^{54}$Fe | 2949.2 →2538.1 | $6^+ \rightarrow 4^+$ | E2 | 100 |
| 783.3 | | $^{50}$Cr | 783.31 →g.s | $2^+ \rightarrow 0^+$ | E2 | 100 |
| 846.76 | | $^{56}$Fe | 846.78 →g.s | $2^+ \rightarrow 0^+$ | E2 | 100 |
| | 847 | $^{55}$Fe | 2255.5 →1408.45 | $\rightarrow 7/2^-$ | | 100 |
| 931.29 | | $^{55}$Fe | 931.29 →g.s | $5/2^- \rightarrow 3/2^-$ | M1+E2 | 98.04 |
| | 931.2 | $^{54}$Mn | 2715.2 →1784.04 | $5^+ \rightarrow 7^+$ | | 100 |
| | 931.04 | $^{54}$Mn | 2856.55 →1925.35 | $8^+ \rightarrow 7^-$ | M1 | 22.76 |
| 935.53 | | $^{52}$Cr | 2369.63 →1434.091 | $4^+ \rightarrow 2^+$ | E2 | 100 |
| | 936.9 | $^{54}$Fe | 6864.3 →5927.4 | $8^+ \rightarrow 7^+$ | M1(+E2) | 48.78 |
| 1098.1 | | $^{50}$Cr | 1881.42 →783.31 | $4^+ \rightarrow 2^+$ | E2 | 100 |
| 1121.7 | | $^{53}$Mn | 2562.99 →1441.15 | $13/2^+ \rightarrow 11/2^-$ | M1+(E2) | 100 |
| 1129.9 | | $^{54}$Fe | 2538.1 →1408.19 | $4^+ \rightarrow 2^+$ | E2 | 100 |
| 1222.5 | | $^{55}$Fe | 2539.11 →1316.54 | $11/2^- \rightarrow 7/2^-$ | E2 | 94.07 |
| 1238.27 | | $^{56}$Fe | 2085.1 →846.78 | $4^+ \rightarrow 2^+$ | E2 | 100 |
| 1253.7 | | $^{52}$Mn | 1253.7 →g.s | $5^+ \rightarrow 6^+$ | | 80.65 |
| | 1251.1 | $^{51}$Mn | 1488.5 →237.3 | $11/2^- \rightarrow 7/2^-$ | E2 | <63.25 |
| | 1252.1 | $^{53}$Mn | 2693.29 →1441.15 | $15/2^- \rightarrow 11/2^-$ | E2 | 94.34 |
| 1282.5 | | $^{50}$Cr | 3164.06 →1881.42 | $6^+ \rightarrow 4^+$ | E2 | 100 |
| 1289.83 | | $^{53}$Mn | 1289.83 →g.s | $3/2^- \rightarrow 7/2^-$ | E2 | 54.05 |
| | 1289.59 | $^{53}$Cr | 1289.52 →g.s | $7/2^- \rightarrow 3/2^-$ | E2 | 94.34 |
| 1303.4 | | $^{56}$Fe | 3388.55 →2085.1 | $6^+ \rightarrow 4^+$ | E2 | 98.72 |
| 1316.4 | | $^{55}$Fe | 1316.54 →g.s | $7/2^- \rightarrow 3/2^-$ | E2 | 92.94 |
| | 1312.6 | $^{56}$Fe | 4683.04 →3369.95 | $(2^+),3^+ \rightarrow 2^+$ | E2 | <19.28 |
| 1328.2 | | $^{53}$Fe | 1328.01→g.s | $9/2^- \rightarrow 7/2^-$ | M1(+E2) | 100 |
| | 1328.95 | $^{52}$Fe | 5654.5 →4325.5 | $6^+ \rightarrow 6^+$ | | 100 |
| 1333.65 | | $^{52}$Cr | 2767.77 →1434.1 | $4^+ \rightarrow 2^+$ | E2 | 98.27 |
| 1369.7 | | $^{55}$Fe | 2301.06 →931.29 | $(9/2) \rightarrow 5/2^-$ | (E2) | <89.93 |
| 1408.4 | | $^{55}$Fe | 1408.45 →g.s | $7/2^- \rightarrow 3/2^-$ | E2 | 44.28 |
| | 1408.1 | $^{54}$Fe | 1408.19 →g.s | $2^+ \rightarrow 0^+$ | E2 | 100 |
| 1434.07 | | $^{52}$Cr | 1434.1→g.s | $2^+ \rightarrow 0^+$ | E2 | 100 |
| 1441.2 | | $^{53}$Mn | 1441.15→g.s | $11/2^- \rightarrow 7/2^-$ | E2 | 100 |
| 1480.3 | | $^{51}$Cr | 1480.07→g.s | $11/2^- \rightarrow 7/2^-$ | E2 | 51.89 |
| 1810.76 | | $^{56}$Fe | 2657.59→846.78 | $2^+ \rightarrow 2^+$ | M1+E2 | 96.99 |
| 2273.2 | | $^{56}$Fe | 3120.11→846.78 | $(1)^+ \rightarrow 2^+$ | | <94.46 |
| | 2273.1 | $^{55}$Mn | 2398.4 →125.95 | $9/2^- \rightarrow 7/2^-$ | | 74.07 |

### 4.2 γ-ray Angular distributions

The experimental γ-ray differential cross sections from the 32 HP-Ge detectors produced mainly in (p, p'γ) inelastic proton scattering were found to be dominated by E2 and (M1 + E2) transitions. They were averaged for each observation angle $\theta$ relative to the incident proton beam direction and versus the γ-ray line energy assuming $\theta_{lab} \approx \theta_{c.m}$. As illustrative examples, we report in Fig. 4 the experimental data versus the proton beam energy for some prominent γ-ray lines produced in proton-induced nuclear reactions on the $^{nat}$Mg, $^{nat}$Si and $^{56}$Fe targets at $E_p$ = 66 and 125 MeV, together with the associated best-fit curves generated by the Legendre polynomial expansion given by Eq. (4). Tables 5 - 7 display the obtained corresponding $a_\ell$ coefficients.

The reduced $\chi^2$-values resulting from the fits amounted to at most 1 for the narrow, isolated γ-ray lines, and to at most 3 in case of overlapping, broader line structures within the γ-ray energy spectra. We derived the total relative uncertainties in the experimental differential cross sections (also shown in Fig. 4) for all the analyzed γ-ray transitions by adding in quadrature the uncertainties in the quantities involved in Eq. (3), i.e., the contributions from the photopeak area determination, the absolute detection efficiency, the target thickness (mean value) amounting to 3%, and the collected beam charge (1%). These uncertainties were combined to obtain the total relative uncertainty for each HP-Ge crystal. The uncertainties in the mean differential cross section values at each observation angle were defined as the absolute differences between the maximum values and the mean values. Finally, an overall, average relative uncertainty of 22 % was obtained. Alternatively, the total relative uncertainties in the experimental angular distributions were estimated by adding in quadrature the uncertainties affecting the parameters of the Legendre polynomial expansion in Eq. (4), leading to consistent values lying within the range of 20-24%.

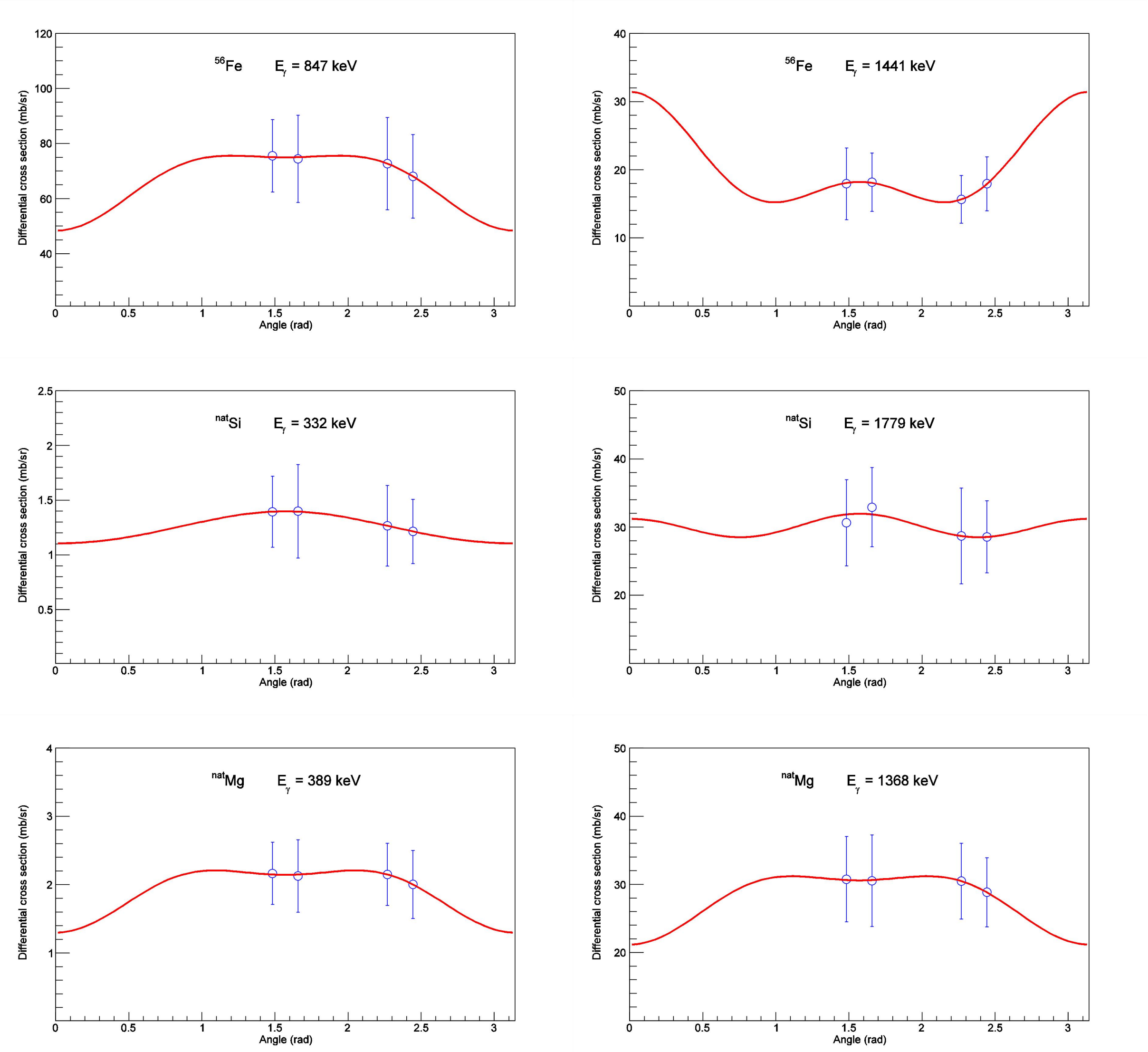

(a) Proton energy $E_p = 66$ MeV. (b) Proton energy $E_p = 125$ MeV.

Fig. 4. Illustrative examples of experimental γ-ray line angular distributions for proton-induced nuclear reactions on the $^{nat}$Mg, $^{nat}$Si and $^{56}$Fe targets (blue symbols) and corresponding Legendre polynomial expansion best fit curves (in red color).

Table 5. List of the Legendre polynomial $a_l$ best fit coefficients obtained by fitting Eq. (4) to the γ-ray angular distributions for the $^{nat}$Mg target.

| γ-ray energy (keV) | $E_p$ = 66 MeV | | | $E_p$ = 80 MeV | | | $E_p$ = 95 MeV | | | $E_p$ = 110 MeV | | | $E_p$ = 125 MeV | | |
|---|---|---|---|---|---|---|---|---|---|---|---|---|---|---|---|
| | $a_0$ | $a_2$ | $a_4$ | $a_0$ | $a_2$ | $a_4$ | $a_0$ | $a_2$ | $a_4$ | $a_0$ | $a_2$ | $a_4$ | $a_0$ | $a_2$ | $a_4$ |
| | | | | | | | **p + $^{nat}$Mg** | | | | | | | | |
| **331.91** | 7.13 | 0.40 | 0.35 | 6.88 | 0.10 | -0.17 | 6.64 | -0.19 | -0.70 | 6.08 | -0.70 | -2.59 | 5.03 | -1.55 | -2.25 |
| **350.73** | 23.23 | -1.13 | 0.83 | 21.78 | -2.97 | -1.17 | 20.33 | -4.81 | -3.17 | 18.44 | -5.47 | -4.04 | 16.96 | -0.15 | -0.77 |
| **389.71** | 2.06 | -0.43 | -0.34 | 1.77 | -0.40 | -0.29 | 1.49 | -0.36 | -0.25 | 1.31 | -0.21 | -0.11 | 1.19 | -0.25 | -0.17 |
| **425.81** | 0.92 | -0.27 | | 0.71 | -0.21 | | 0.51 | -0.14 | | 0.36 | -0.16 | | 0.30 | 0.0 | |
| **439.99** | 35.55 | -4.05 | -2.57 | 33.39 | -4.66 | -4.02 | 31.22 | -5.28 | -5.46 | 29.39 | -2.21 | -2.31 | 27.15 | 13.59 | 1.73 |
| **450.7** | 37.01 | -5.15 | -3.90 | 35.89 | -5.58 | -4.89 | 34.77 | -6.00 | -5.88 | 31.24 | -6.08 | -5.61 | 27.69 | 1.74 | -1.43 |
| **472.2** | 4.06 | -0.22 | | 3.63 | -0.06 | | 3.21 | 0.11 | | 3.20 | 0.05 | | 2.73 | -0.39 | |
| **563.19** | 0.85 | 0.10 | 0.47 | 0.79 | -0.07 | -0.43 | 0.73 | -0.04 | 0.39 | 0.76 | -0.58 | -0.24 | 0.69 | 0.01 | 0.22 |
| **585.03** | 34.22 | -7.70 | -7.95 | 31.42 | -5.77 | -6.59 | 28.62 | -3.84 | -5.23 | 26.04 | -6.63 | -6.57 | 23.38 | -0.34 | -7.99 |
| **937.0** | 7.36 | 1.77 | 1.08 | 7.02 | 0.54 | 0.13 | 6.68 | -0.69 | -0.81 | 6.46 | 0.04 | 0.32 | 5.92 | -0.07 | -1.10 |
| **974.74** | 2.20 | -0.51 | -0.08 | 1.79 | -0.21 | 0.02 | 1.39 | 0.10 | 0.11 | 1.28 | 0.21 | 0.22 | 1.04 | 0.09 | 0.01 |
| **1016.9** | 2.77 | 0.37 | 0.88 | 2.44 | -0.15 | 0.22 | 2.11 | -0.67 | -0.45 | 2.05 | -0.68 | -0.68 | 1.66 | -0.72 | -0.48 |
| **1368.63** | 56.53 | -2.96 | -1.68 | 49.22 | -2.93 | -0.57 | 41.90 | -2.90 | 0.54 | 34.60 | -1.59 | 0.55 | 29.53 | -4.83 | -3.60 |
| **1808.68** | 5.28 | -0.31 | 0.36 | 4.57 | -1.33 | 0.13 | 3.86 | -2.35 | -0.10 | 3.42 | -0.05 | 0.63 | 2.76 | -1.44 | 0.27 |
| **2754.02** | 12.05 | 4.85 | 8.25 | 11.09 | 1.28 | 5.23 | 10.12 | -2.30 | 2.20 | 9.66 | 1.89 | 5.25 | 9.44 | 1.82 | 3.26 |

Table 6. Same as in Table 5 but for the $^{nat}$Si target.

| γ-ray energy (keV) | $E_p$ = 66 MeV | | | $E_p$ = 80 MeV | | | $E_p$ = 95 MeV | | | $E_p$ = 110 MeV | | | $E_p$ = 125 MeV | | |
|---|---|---|---|---|---|---|---|---|---|---|---|---|---|---|---|
| | $a_0$ | $a_2$ | $a_4$ | $a_0$ | $a_2$ | $a_4$ | $a_0$ | $a_2$ | $a_4$ | $a_0$ | $a_2$ | $a_4$ | $a_0$ | $a_2$ | $a_4$ |
| | | | | | | | **p + $^{nat}$Si** | | | | | | | | |
| **331.91** | 1.29 | -0.20 | 0.01 | 1.99 | -0.02 | -0.39 | 2.13 | -1.26 | -1.02 | 2.32 | -0.22 | -0.11 | 2.19 | -0.55 | -0.28 |
| **350.73** | 6.59 | -1.35 | -0.78 | 9.96 | -2.43 | -3.19 | 10.98 | -2.58 | -3.19 | 10.21 | -2.64 | -2.37 | 10.59 | -3.16 | -3.36 |
| **389.71** | 5.08 | 0.05 | -0.52 | 4.86 | -1.24 | -1.46 | 4.49 | -1.19 | -1.32 | 3.70 | -0.80 | -1.84 | 3.58 | -1.61 | -1.83 |
| **416.85** | 30.54 | -6.97 | -7.35 | 29.28 | -3.61 | -5.67 | 25.03 | -6.01 | -10.18 | 22.42 | -5.95 | -7.52 | 19.58 | -3.68 | -7.10 |
| **439.99** | 16.86 | -4.85 | -4.86 | 13.64 | -7.02 | -8.16 | 15.14 | -5.31 | -3.91 | 13.77 | -5.48 | -5.85 | 13.90 | -4.52 | 4.76 |
| **450.7** | 10.32 | -2.87 | -3.83 | 8.71 | -2.50 | -4.59 | 8.36 | -3.03 | -4.29 | 7.69 | -3.46 | -3.91 | 6.42 | -2.17 | -4.04 |
| **585.03** | 21.14 | -3.50 | -2.85 | 21.37 | -2.04 | -2.79 | 20.64 | -3.89 | -4.03 | 18.77 | 0.45 | -2.64 | 18.44 | -0.20 | -0.58 |
| **780.8** | 14.57 | -3.31 | -3.23 | 13.88 | -4.75 | -6.03 | 13.81 | -2.26 | -2.35 | 9.25 | -2.66 | -2.68 | 10.92 | -0.72 | -0.75 |
| **843.76** | 10.01 | -1.95 | -1.57 | 9.12 | -3.62 | -3.24 | 8.82 | -1.89 | -1.45 | 7.51 | -3.58 | -3.95 | 7.02 | -2.64 | -3.09 |
| **890.87** | 3.81 | -0.22 | -0.37 | 3.58 | -1.18 | -1.56 | 3.54 | -0.43 | -0.21 | 3.87 | -1.13 | -1.05 | 3.41 | -3.60 | -3.02 |
| **957.3** | 13.00 | -0.77 | -1.09 | 12.25 | -1.45 | -2.00 | 11.30 | -1.60 | -1.77 | 11.08 | -3.18 | -3.25 | 8.86 | -0.65 | -0.68 |
| **974.74** | 4.82 | -0.65 | -0.37 | 5.00 | 0.09 | -0.19 | 4.66 | -1.29 | -1.27 | 4.30 | -0.85 | -0.52 | 3.78 | -1.36 | -1.30 |
| **1273.36** | 7.83 | -1.98 | -1.61 | 7.21 | -2.03 | -2.63 | 6.92 | -0.46 | -0.72 | 6.69 | -0.19 | -0.33 | 6.57 | -0.94 | -0.50 |
| **1368.63** | 44.03 | -0.41 | 0.54 | 48.09 | -0.08 | 0.42 | 45.59 | -4.52 | -1.65 | 41.94 | -5.44 | -1.44 | 43.58 | 6.19 | 5.45 |
| **1778.97** | 61.85 | -8.64 | -7.05 | 52.81 | -5.06 | -1.48 | 43.70 | -5.76 | -3.02 | 34.86 | -0.756 | 1.52 | 30.08 | -1.66 | 2.84 |

Table 7. Same as in Table 5 but for the $^{56}Fe$ target.

| γ-ray energy (keV) | $E_p$ = 66 MeV | | | $E_p$ = 80 MeV | | | $E_p$ = 95 MeV | | | $E_p$ = 110 MeV | | | $E_p$ = 125 MeV | | |
|---|---|---|---|---|---|---|---|---|---|---|---|---|---|---|---|
| | $a_0$ | $a_2$ | $a_4$ | $a_0$ | $a_2$ | $a_4$ | $a_0$ | $a_2$ | $a_4$ | $a_0$ | $a_2$ | $a_4$ | $a_0$ | $a_2$ | $a_4$ |
| | | | | | | | **p + $^{56}$Fe** | | | | | | | | |
| **125.95** | 12.46 | 0.08 | 0.67 | 13.62 | -3.79 | -1.76 | 12.83 | -1.33 | 0.44 | 11.87 | -0.54 | 1.49 | 9.15 | 5.43 | 4.11 |
| **156.29** | 54.10 | -8.49 | 0.85 | 52.89 | -5.84 | 4.88 | 45.92 | -7.12 | -0.51 | 41.31 | -7.45 | -2.55 | 33.66 | 8.25 | 7.65 |
| **211.98** | 28.89 | -2.55 | | 28 | -3.3 | | 23.81 | -2.33 | | 19.7 | -0.94 | | 16.16 | -0.77 | |
| **226.3** | 10.84 | -1.50 | -1.10 | 9.85 | -0.49 | -0.08 | 10.27 | -2.01 | -1.88 | 11.08 | -0.79 | -0.64 | 8.37 | -1.30 | -0.26 |
| **274.8** | 8.99 | -1.73 | -1.54 | 13.54 | -1.61 | 0.36 | 10.76 | -1.31 | -0.27 | 8.72 | -2.67 | -1.28 | 7.40 | 1.83 | 1.11 |
| **377.88** | 35.45 | -11.86 | -9.62 | 33.75 | -9.07 | -7.94 | 28.77 | -8.90 | -7.22 | 25.13 | -7.31 | -5.83 | 22.87 | -6.65 | -5.30 |
| **411.9** | 31.72 | -5.10 | -8.19 | 29.61 | -7.72 | -9.12 | 24.68 | -9.82 | -6.81 | 21.02 | -2.50 | -1.63 | 17.96 | -2.14 | -1.39 |
| **783.3** | 33.4 | -5.73 | -7.65 | 25.72 | -5.24 | -5.46 | 26.62 | -5.19 | -7.72 | 29.06 | -1.49 | -1.71 | 20.82 | -2.08 | -3.09 |
| **846.77** | 70.90 | -14.43 | -8.32 | 64.68 | -11.01 | -7.47 | 49.33 | -18.3 | -11.81 | 38.72 | -12.82 | -8.02 | 29.82 | 2.64 | 0.83 |
| **931.29** | 44.68 | -3.46 | -4.04 | 41.4 | -6.28 | -12.47 | 33.85 | -20.34 | -14.88 | 28.39 | -9.60 | -7.86 | 22.19 | -8.62 | -9.58 |
| **935.53** | 18.36 | 0.82 | -4.19 | 23.32 | 1.90 | -1.76 | 18.95 | -3.91 | -7.44 | 16.56 | -2.89 | -4.77 | 14.57 | -11.04 | -13.33 |
| **1098.1** | 23.71 | -2.81 | -2.81 | 17.52 | -1.07 | -1.00 | 19.16 | -2.49 | -2.21 | 20.64 | -2.28 | -1.26 | 15.78 | 3.39 | 4.37 |
| **1121.7** | 11.66 | -4.87 | -3.18 | 12.98 | -4.23 | -2.84 | 11.61 | -1.10 | -0.68 | 10.47 | -1.38 | -1.73 | 10.42 | -4.28 | -9.17 |
| **1129.9** | 38.55 | -1.37 | -1.53 | 34.22 | -2.75 | -0.53 | 27.38 | -2.61 | -1.28 | 22.08 | -2.08 | -0.31 | 22.22 | -6.81 | -7.15 |
| **1222.5** | 14.69 | 0.26 | 0.31 | 13.28 | -2.56 | -0.70 | 11 | -2.57 | -1.15 | 9.42 | -1.18 | 0.12 | 8.48 | -0.73 | -0.17 |
| **1238.27** | 20.39 | -4.70 | -3.42 | 19.08 | -3.38 | -0.17 | 15.51 | -4.27 | -1.55 | 11.89 | -4.11 | -2.72 | 10.21 | -6.02 | -3.79 |
| **1253.7** | 16.65 | 2.53 | -1.13 | 13.11 | -2.82 | -3.23 | 11.63 | -0.70 | -1.03 | 9.62 | -0.33 | -0.26 | 7.51 | 2.5 | 0.73 |
| **1282.5** | 12.40 | -1.92 | -2.64 | 9.55 | 0.48 | 0.54 | 9.93 | -1.59 | -1.64 | 12.26 | -2.36 | -1.84 | 9.98 | 3.36 | 3.15 |
| **1289.83** | 11.11 | 0.51 | 1.65 | 11.78 | 0.23 | 0.73 | 10.21 | -2.85 | -1.97 | 9.11 | -2.37 | -1.09 | 8.14 | -2.11 | -0.97 |
| **1303.4** | 3.48 | 0.91 | 0.54 | 3.56 | 1.72 | 2.23 | 2.53 | -0.71 | -0.51 | 2.14 | -1.28 | -0.72 | 1.93 | -6.04 | -5.29 |
| **1316.04** | 38.27 | 2.67 | 1.39 | 35.41 | -5.07 | -2.42 | 32.49 | -3.75 | -2.74 | 27.55 | -3.73 | -2.16 | 21.00 | 5.76 | 6.93 |
| **1328.2** | 3.04 | -1.83 | -1.64 | 2.70 | -3.44 | -3.97 | 2.30 | -1.59 | -1.76 | 2.17 | 0.01 | 0.81 | 1.96 | 1.97 | 3.41 |
| **1333.65** | 11.51 | -4.37 | -2.71 | 15.35 | -0.57 | -0.14 | 13.13 | 0.67 | 1.08 | 11.72 | 0.27 | 1.35 | 10.46 | 0.24 | 1.20 |
| **1369.7** | 7.05 | 2.59 | 2.95 | 6.49 | -2.64 | -1.42 | 5.47 | -0.33 | -1.67 | 4.80 | 3.64 | 1.60 | | | |
| **1408.4** | 87.04 | -14.92 | -16.15 | 84.34 | -5.92 | -0.33 | 72.29 | 0.78 | 5.11 | 57.82 | -2.79 | 0.12 | 45.56 | 4.28 | 7.67 |
| **1434.07** | 54.21 | -8.47 | -7.74 | 64.29 | -4.15 | -5.68 | 60.36 | -1.93 | 0.36 | 51.69 | -6.18 | -3.38 | 42.71 | 8.31 | 12.53 |
| **1441.2** | 47.79 | 3.38 | -0.20 | 36.53 | 0.62 | -1.46 | 30.17 | 0.04 | 2.21 | 25.04 | -1.21 | 3.31 | 18.2 | 5.67 | 7.69 |
| **1480.3** | 4.01 | -0.89 | -1.05 | 7.19 | -2.54 | -1.95 | 9.28 | 0.58 | 1.24 | 7.81 | 0.08 | 0.10 | 5.89 | 1.86 | 2.28 |
| **1810.76** | 8.81 | 2.74 | 4.04 | 7.01 | -0.04 | 2.16 | 6.57 | -3.55 | 0.84 | 5.17 | -2.59 | -1.43 | 4.85 | -3.32 | -0.76 |
| **2273.2** | 1.79 | -1.09 | -0.88 | 1.47 | -2.19 | 2.18 | 1.15 | -1.55 | -0.86 | 0.93 | -0.35 | -0.49 | 0.77 | -0.43 | -0.70 |

### 4.3 Total γ-ray line production cross sections

In order to ensure the homogeneity and consistency of the cross sections derived from different experimental runs, particularly those performed with targets of different isotopic compositions, we normalized the cross section experimental data sets from Refs. [17, 24, 25, 29] involving the $^{nat}Fe$ target to the isotopic composition of $^{56}Fe$. Similarly, the cross section data sets from Refs. [20, 24, 25] taken with the $^{24}Mg$ target were adjusted to the natural isotopic abundance of magnesium.

The obtained σ ($E_p$) experimental results are reported in Figs. (5 - 12) for all the analyzed γ-ray lines, where they are compared to previous experimental data sets from the literature when available. They concern a total of sixty γ-ray lines produced in nuclear reactions induced by 66, 80, 100, 110 and 125 MeV proton beams on the $^{nat}Mg$, $^{nat}Si$ and $^{56}Fe$ targets. More precisely, Figs (5, 6) report the σ($E_p$) results for 15 lines observed in proton irradiation of the $^{nat}Mg$ target, Figs. (7, 8) report similar results for 15 lines from the $^{nat}Si$ target, and Figs. (9 -

12) report the obtained results for 30 γ-ray lines from the $^{56}$Fe target. The numerical values of the total γ-ray line production cross sections from these 3 targets are respectively listed in Tables (8, 9, 10) along with the corresponding average relative uncertainties. The uncertainties in the differential cross sections affect the Legendre polynomial expansion fits to the experimental data and influence the extracted values of the $a_\ell$ coefficients and the associated uncertainties. This effect is particularly significant for the zeroth order term, $a_0$, used for calculating the total γ-ray line cross sections and the associated relative uncertainties, found to amount to 12-19%.

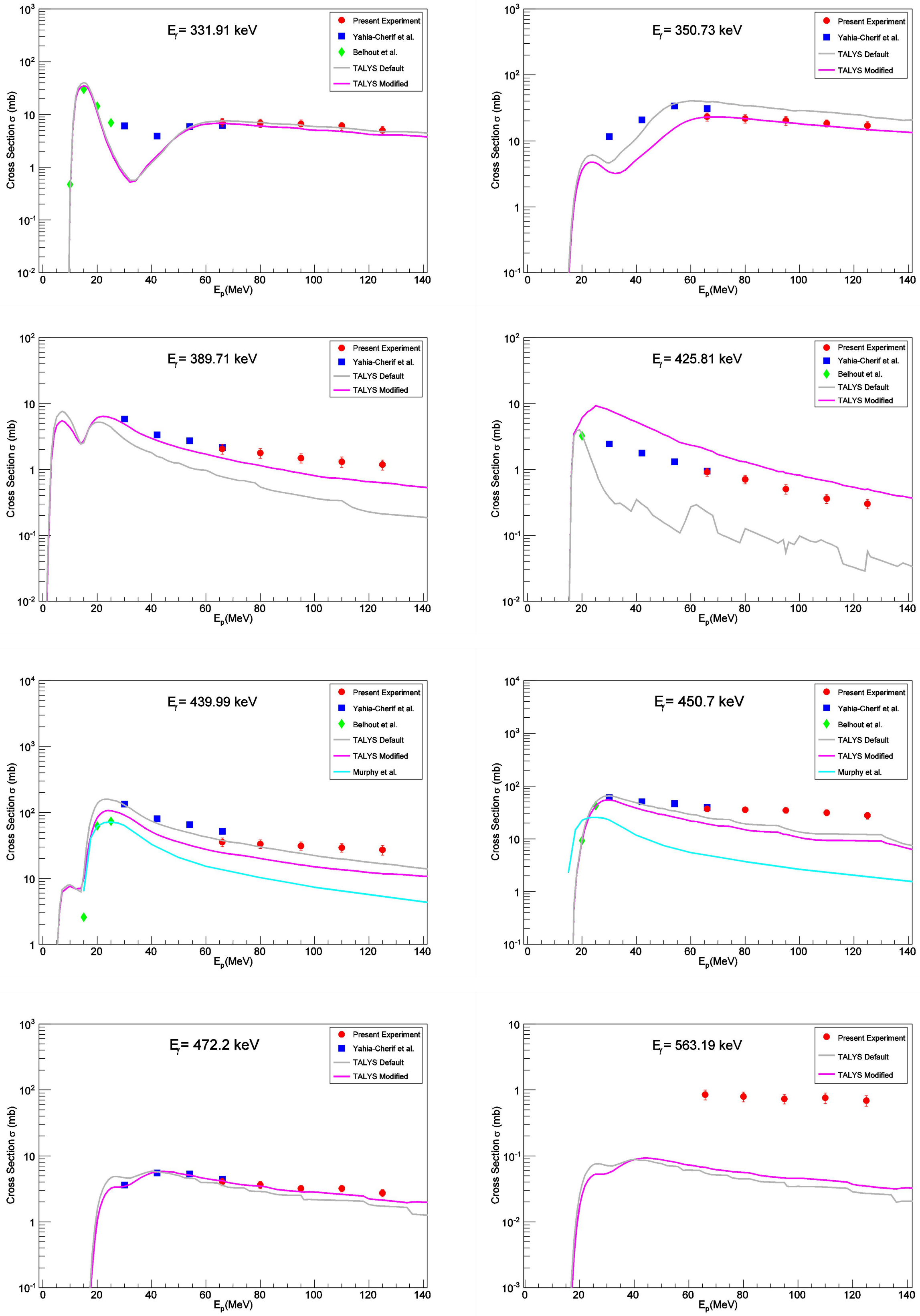

Fig. 5. Total production cross sections for 8 γ-ray lines produced in proton reactions with the $^{nat}Mg$ target. The experimental data are shown as follows: in red full circles (data from this work), in blue full squares (data of Yahia-Cherif et al. [18]), in green full diamonds (data of Belhout et al. [24, 25]), in black, down-oriented full triangles (data of Dyer et al. [20]), and in yellow, up-oriented full triangles (Lesko et al. [29] data). The Teal curves correspond to the predictions of the Murphy et al. semi-empirical compilation [31]. The grey curves represent the results derived by TALYS code [32] calculations with default input parameters, while the pink curves refer to TALYS code calculations using our modified OMP and $\beta_\lambda$ level deformation parameters.

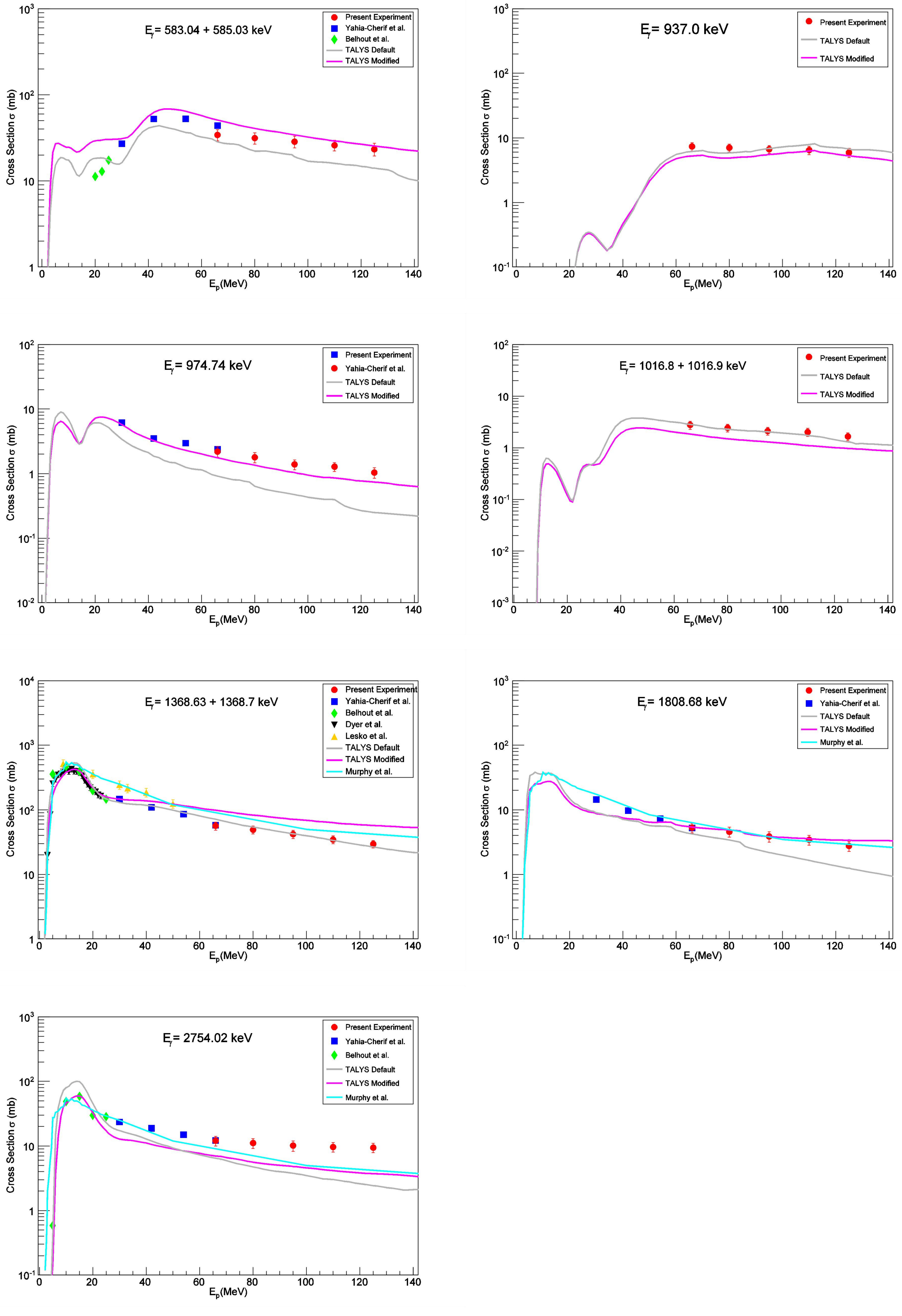

Fig. 6. Same as in Fig.5 but for 7 other γ-ray lines emitted in proton interactions with the $^{nat}$Mg target.

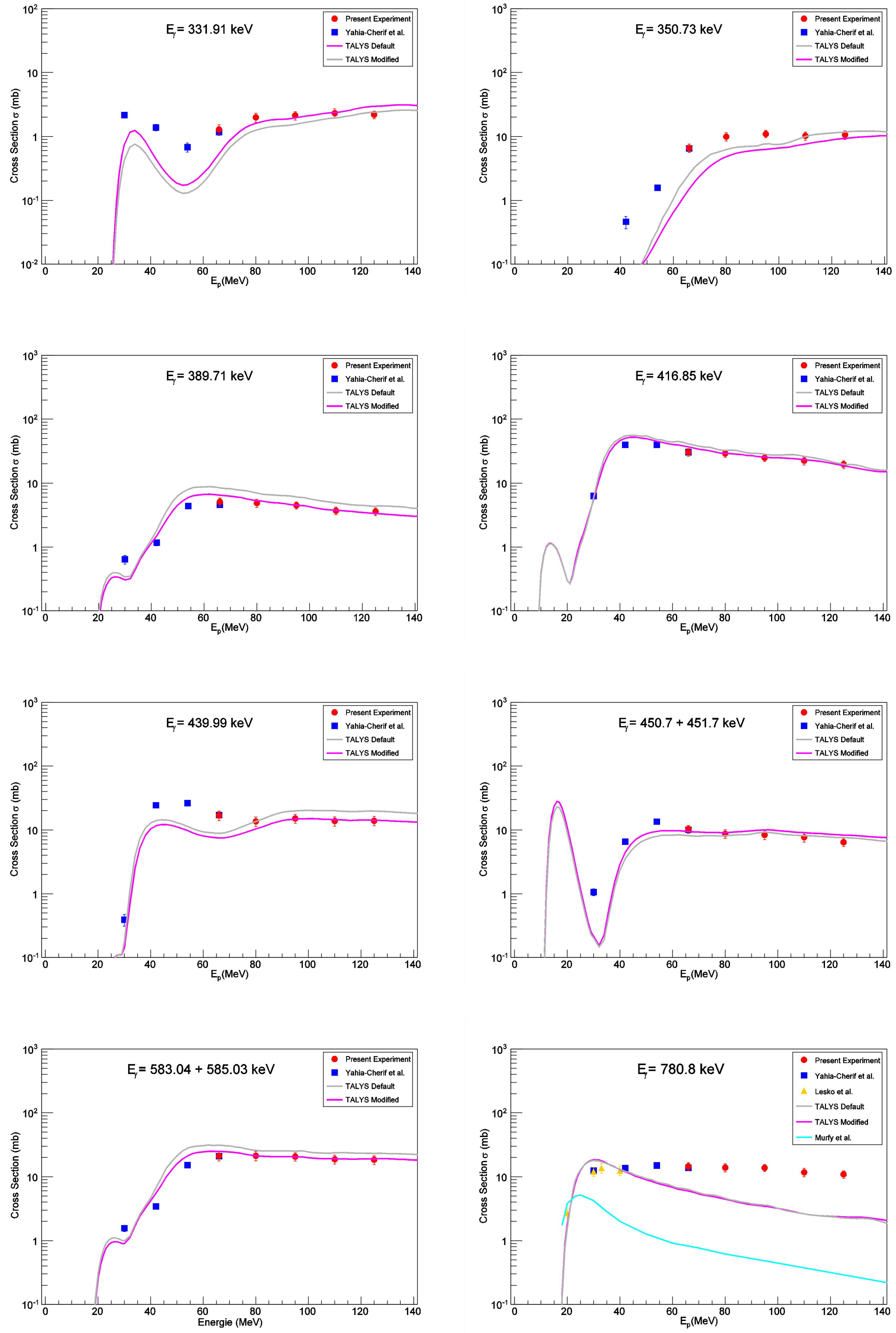

Fig. 7. Same as in Fig. 5 but for 8 γ-ray lines produced in proton reactions with the $^{nat}Si$ target.

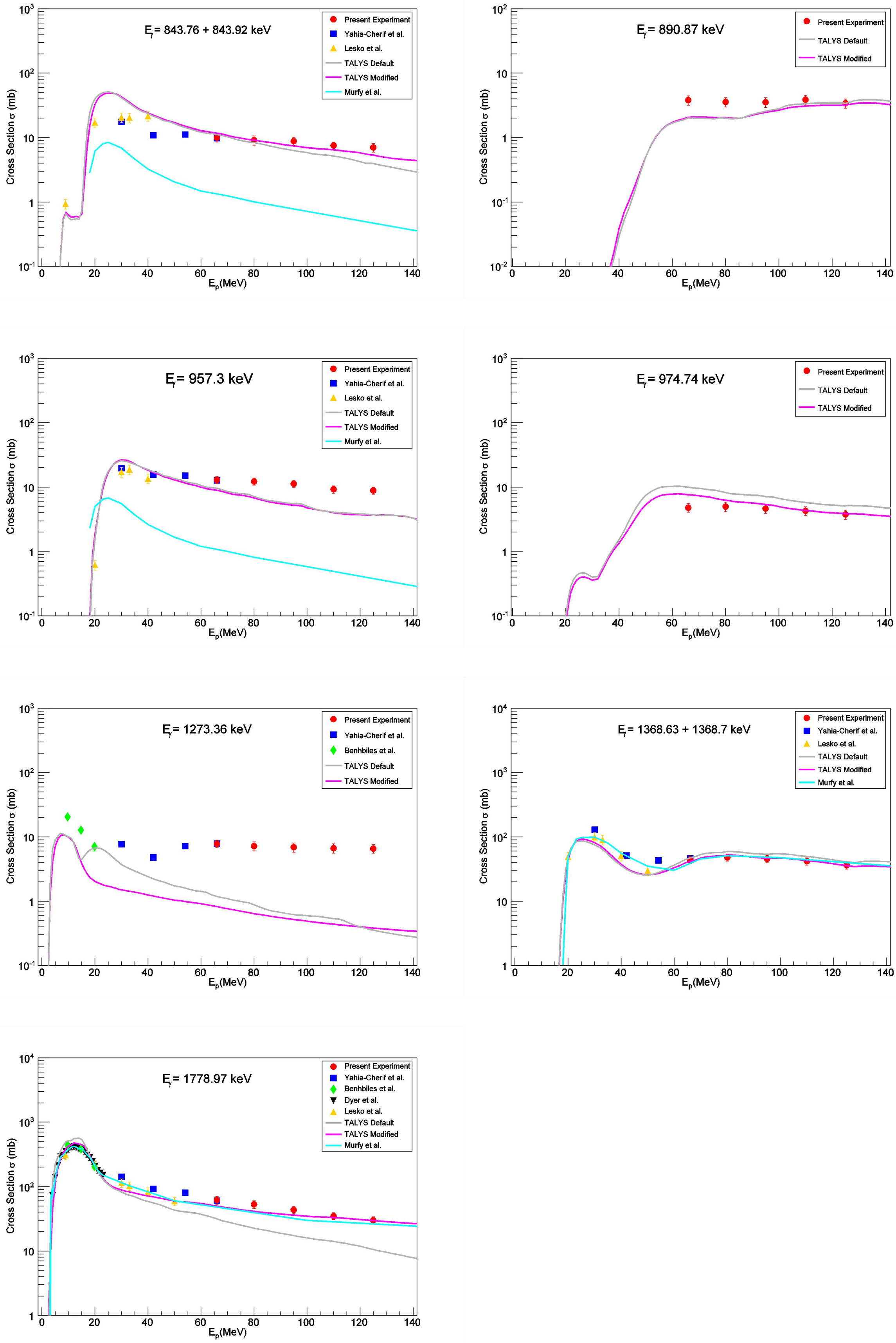

Fig. 8. Same as in Figs. 5 but for seven 7 other γ-ray lines produced in proton reactions with the $^{nat}Si$ target. The same symbols are used, except that here the green circles represent the experimental data of Benhabiles et al. [17].

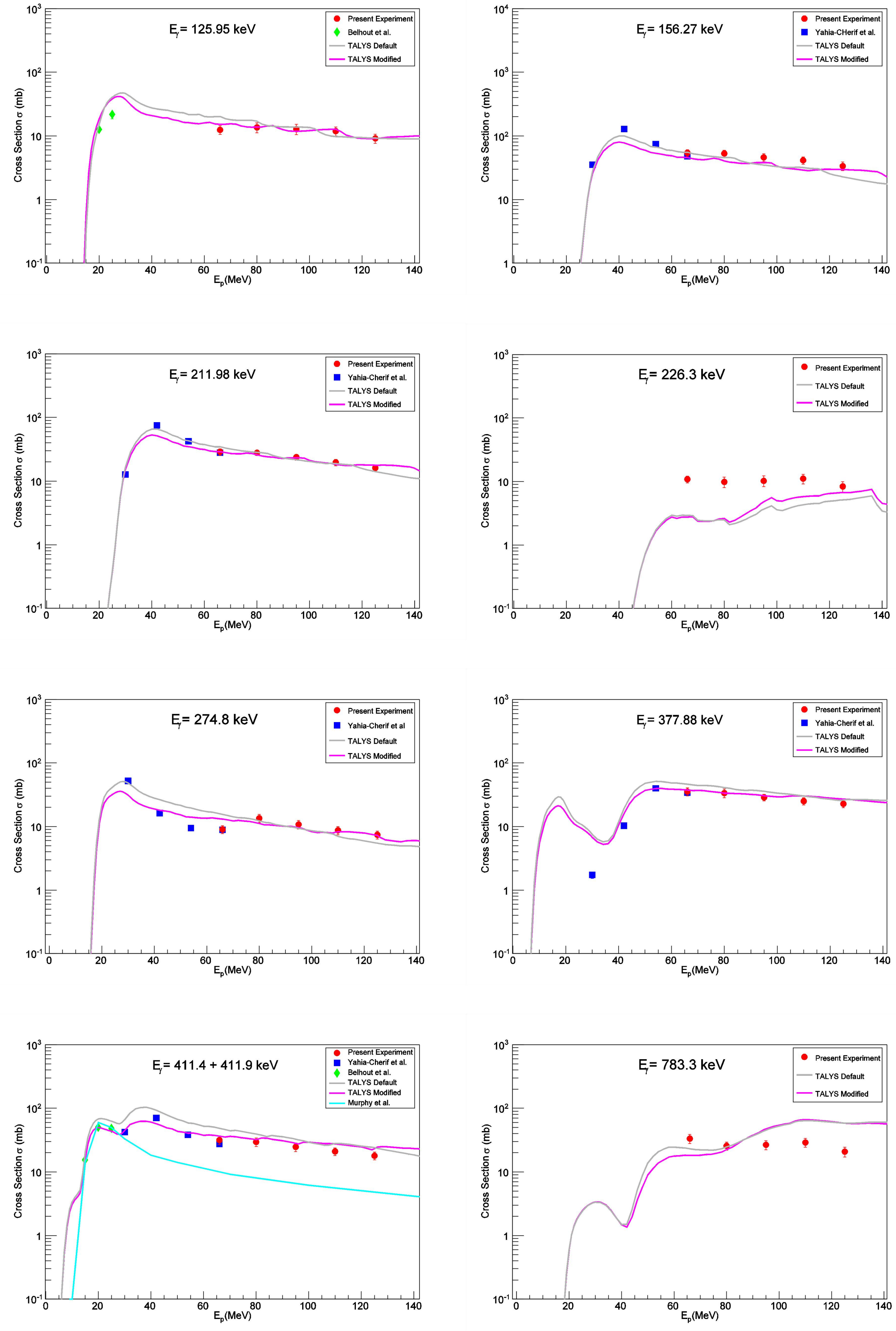

Fig. 9. Same as in Fig. 5 but for 8 γ-ray lines produced in proton reactions with the $^{56}$Fe target.

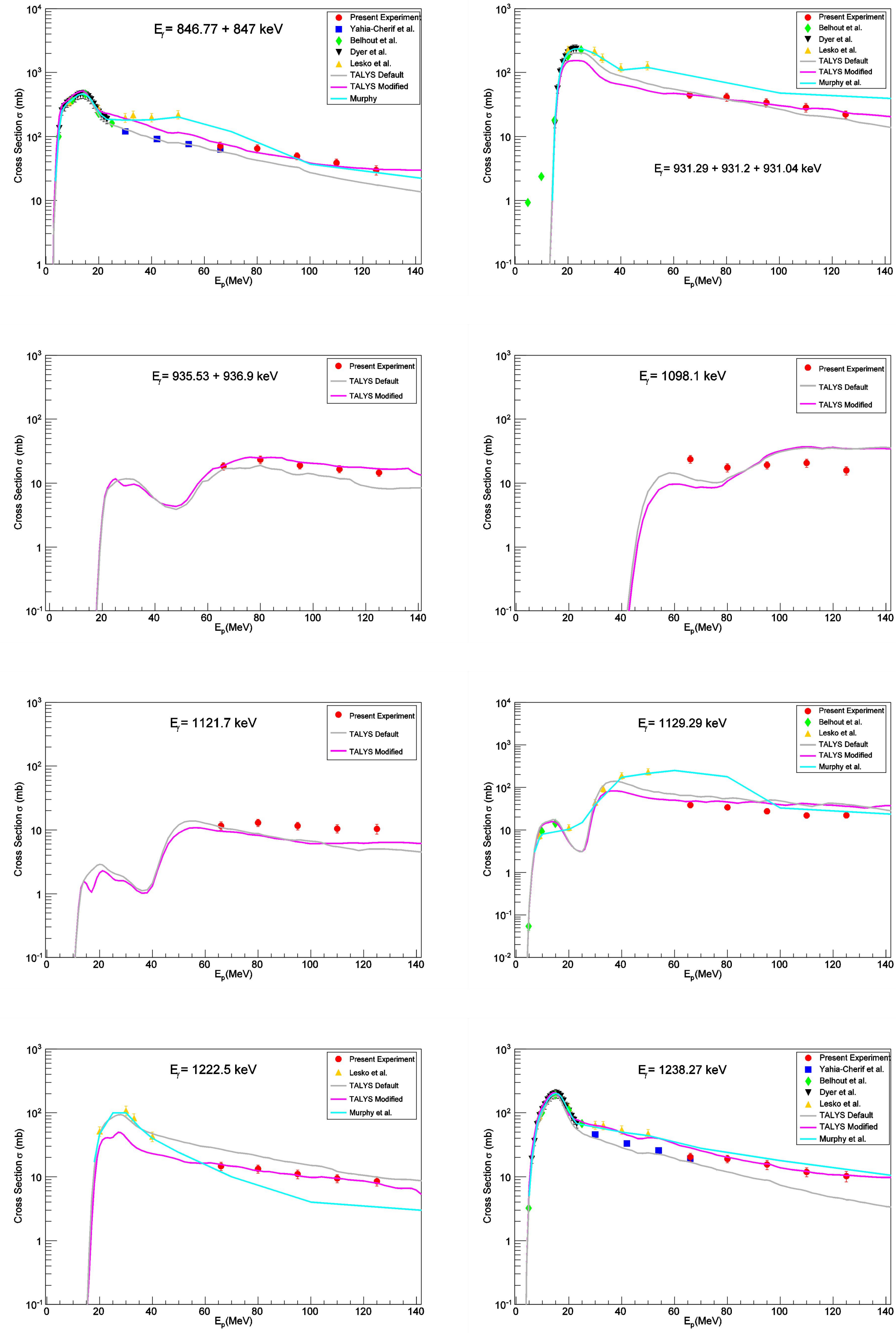

Fig. 10. Same as in Fig.5 but for 8 other γ-ray lines emitted in proton interactions with the $^{56}Fe$ target

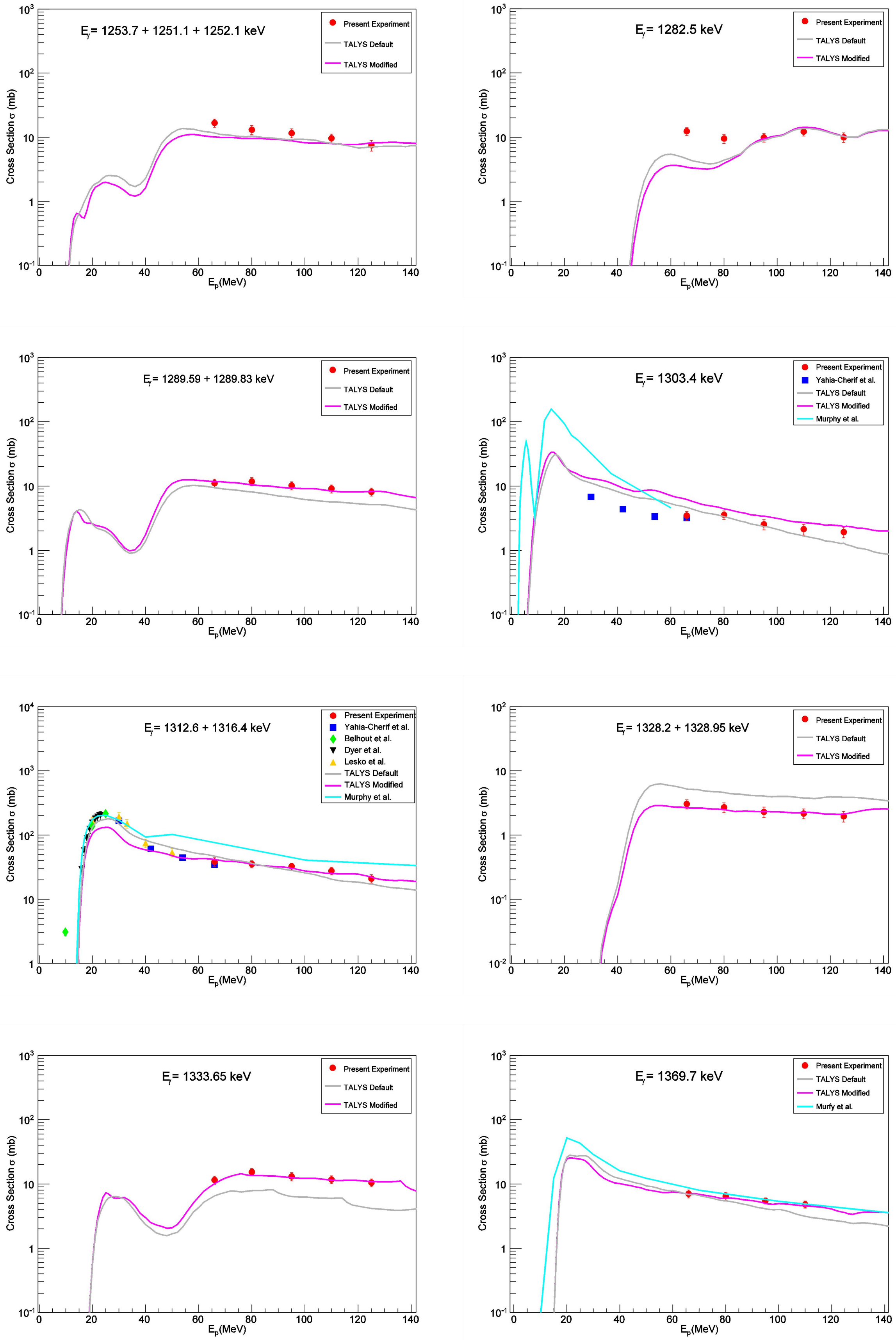

Fig.11. Same as in Fig.5 but for 8 other γ-ray lines emitted in proton interactions with the $^{56}$Fe target.

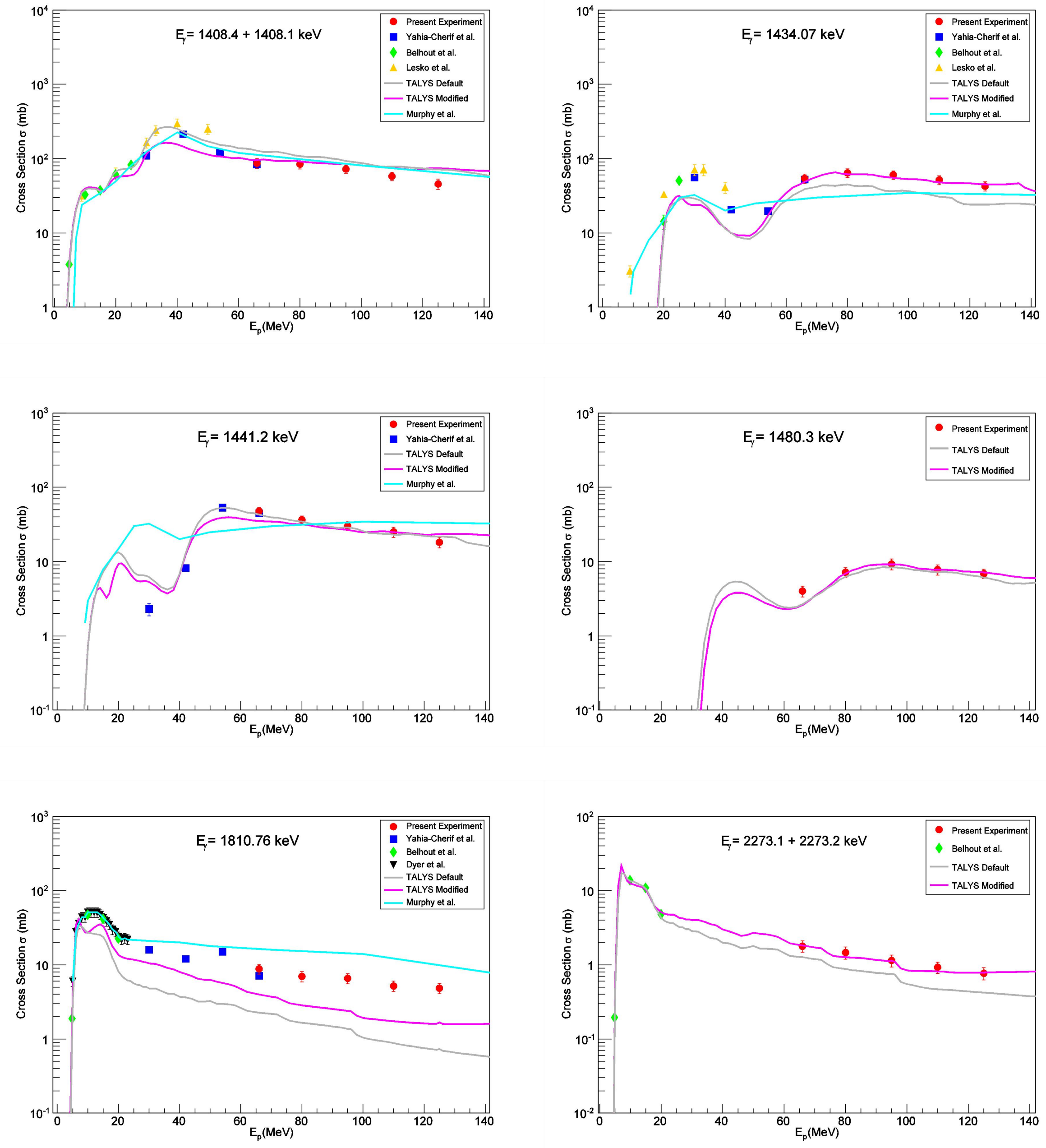

Fig. 12. Same as in Fig.5 but for 6 other γ-ray lines emitted in proton interactions with the $^{56}Fe$ target.

Table 8. Experimental values of total production cross sections and their uncertainties for γ-ray lines emitted in proton irradiations with the $^{nat}Mg$ target

| γ -ray energy (keV) | $E_p$ = 66 MeV | $E_p$ = 80 MeV | $E_p$ = 95 MeV | $E_p$ = 110 MeV | $E_p$ = 125 MeV |
|---|---|---|---|---|---|
| **331.91** | 7.13 ± 1.25 | 6.88 ± 1.23 | 6.64 ± 1.21 | 6.08 ± 1.07 | 5.03 ± 0.93 |
| **350.73** | 23.23 ± 3.26 | 21.78 ± 3.21 | 20.33 ± 3.16 | 18.44 ± 2.25 | 16.96 ± 2.56 |
| **389.71** | 2.06 ± 0.35 | 1.77 ± 0.30 | 1.49 ± 0.24 | 1.31 ± 0.23 | 1.19 ± 0.21 |
| **425.81** | 0.92 ± 0.13 | 0.71 ± 0.11 | 0.51 ± 0.08 | 0.36 ± 0.06 | 0.30 ± 0.05 |
| **439.99** | 35.55 ± 5.36 | 33.39 ± 4.86 | 31.22 ± 4.35 | 29.39 ± 4.36 | 27.15 ± 4.52 |
| **450.7** | 37.01 ± 4.56 | 35.89 ± 4.49 | 34.77 ± 4.42 | 31.24 ± 4.05 | 27.69 ± 4.05 |
| **472.2** | 4.06 ± 0.55 | 3.63 ± 0.46 | 3.21 ± 0.37 | 3.20 ± 0.35 | 2.73 ± 0.33 |
| **563.19** | 0.85 ± 0.14 | 0.79 ± 0.12 | 0.73 ± 0.12 | 0.76 ± 0.14 | 0.69 ± 0.13 |
| **585.03** | 34.22 ± 5.05 | 31.42 ± 4.76 | 28.62 ± 4.40 | 26.04 ± 3.74 | 23.38 ± 3.90 |
| **937.0** | 7.36 ± 1.06 | 7.02 ± 0.99 | 6.68 ± 0.92 | 6.46 ± 0.95 | 5.92 ± 0.97 |
| **974.74** | 2.20 ± 0.41 | 1.79 ± 0.33 | 1.39 ± 0.24 | 1.28 ± 1.99 | 1.04 ± 0.19 |
| **1016.9** | 2.77 ± 0.49 | 2.44 ± 0.46 | 2.11 ± 0.36 | 2.02 ± 0.37 | 1.66 ± 0.29 |
| **1368.63** | 56.53 ± 8.03 | 49.22 ± 7.03 | 41.90 ± 6.03 | 34.60 ± 4.81 | 29.53 ± 3.86 |
| **1808.68** | 5.28 ± 0.90 | 4.57 ± 0.80 | 3.86 ± 0.69 | 3.42 ± 0.59 | 2.76 ± 0.49 |
| **2754.02** | 12.05 ± 2.02 | 11.09 ± 1.94 | 10.12 ± 1.87 | 9.66 ± 1.63 | 9.44 ± 1.58 |

Table 9. Same as in Table 8 for γ-ray lines emitted in proton irradiations with the $^{nat}Si$ target.

| γ -ray energy (keV) | $E_p$ = 66 MeV | $E_p$ = 80 MeV | $E_p$ = 95 MeV | $E_p$ = 110 MeV | $E_p$ = 125 MeV |
|---|---|---|---|---|---|
| **331.91** | 1.29 ± 0.22 | 1.99 ± 0.32 | 2.13 ± 0.32 | 2.32 ± 0.40 | 2.19 ± 0.30 |
| **350.73** | 6.59 ± 0.99 | 9.96 ± 1.46 | 10.98 ± 1.42 | 10.21 ± 1.53 | 10.59 ± 1.63 |
| **389.71** | 5.08 ± 0.74 | 4.86 ± 0.73 | 4.49 ± 0.57 | 3.70 ± 0.49 | 3.58 ± 0.49 |
| **416.85** | 30.54 ± 4.25 | 29.28 ± 3.92 | 25.03 ± 3.19 | 22.42 ± 3.27 | 19.58 ± 2.68 |
| **439.99** | 16.86 ± 2.87 | 13.64 ± 2.28 | 15.14 ± 2.47 | 13.77 ± 2.30 | 13.90 ± 2.32 |
| **450.7** | 10.32 ± 1.49 | 8.71 ± 1.35 | 8.36 ± 1.33 | 7.69 ± 1.28 | 6.42 ± 0.94 |
| **585.03** | 21.14 ± 3.42 | 21.37 ± 3.56 | 20.64 ± 3.04 | 18.77 ± 2.99 | 18.44 ± 2.88 |
| **780.8** | 14.57 ± 2.08 | 13.88 ± 1.99 | 13.81 ± 1.72 | 9.25 ± 1.18 | 10.92 ± 1.46 |
| **843.76** | 10.01 ± 1.46 | 9.12 ± 1.46 | 8.82 ± 1.24 | 7.51 ± 0.97 | 7.02 ± 1.11 |
| **890.87** | 3.81 ± 0.64 | 3.58 ± 0.59 | 3.54 ± 0.63 | 3.87 ± 0.67 | 3.41 ± 0.56 |
| **957.3** | 13.00 ± 1.59 | 12.25 ± 1.58 | 11.30 ± 1.37 | 11.08 ± 1.44 | 8.86 ± 1.12 |
| **974.74** | 4.82 ± 0.74 | 5.00 ± 0.83 | 4.66 ± 0.75 | 4.30 ± 0.67 | 3.78 ± 0.61 |
| **1273.36** | 7.83 ± 1.05 | 7.21 ± 1.15 | 6.92 ± 1.11 | 6.69 ± 1.13 | 6.57 ± 1.00 |
| **1368.63** | 44.03 ± 5.84 | 48.09 ± 6.05 | 45.59 ± 5.74 | 41.94 ± 5.46 | 43.58 ± 5.76 |
| **1778.97** | 61.85 ± 8.73 | 52.81 ± 7.32 | 43.70 ± 5.25 | 34.86 ± 4.30 | 30.08 ± 3.91 |

Table 10. Same as in Table 8 for γ-ray lines emitted in proton irradiations with the $^{56}Fe$ target

| γ -ray energy (keV) | $E_p$ = 66 MeV | $E_p$ = 80 MeV | $E_p$ = 95 MeV | $E_p$ = 110 MeV | $E_p$ = 125 MeV |
|---|---|---|---|---|---|
| **125.95** | 12.46 ± 1.95 | 13.62 ± 2.41 | 12.83 ± 2.40 | 11.87 ± 1.90 | 9.15 ± 1.49 |
| **156.29** | 54.10 ± 6.57 | 52.89 ± 6.48 | 45.92 ± 6.61 | 41.31 ± 5.43 | 33.66 ± 5.24 |
| **211.98** | 28.89 ± 3.35 | 28 ± 2.75 | 23.81 ± 2.54 | 19.7 ± 2.27 | 16.16 ± 1.92 |
| **226.3** | 10.84 ± 1.31 | 9.85 ± 1.85 | 10.27 ± 1.97 | 11.08 ± 1.93 | 8.37 ± 1.60 |
| **274.8** | 8.99 ± 1.29 | 13.54 ± 1.97 | 10.76 ± 1.65 | 8.72 ± 1.28 | 7.40 ± 1.08 |
| **377.88** | 35.45 ± 5.09 | 33.75 ± 5.33 | 28.77 ± 3.47 | 25.13 ± 3.28 | 22.87 ± 3.10 |
| **411.9** | 31.72 ± 4.79 | 29.61 ± 4.49 | 24.68 ± 3.80 | 21.02 ± 2.88 | 17.96 ± 2.56 |
| **783.3** | 33.4 ± 5.42 | 25.72 ± 3.79 | 26.62 ± 4.28 | 29.06 ± 4.67 | 20.82 ± 3.62 |
| **846.77** | 70.90 ± 10.69 | 64.68 ± 9.54 | 49.33 ± 5.94 | 38.72 ± 5.53 | 29.82 ± 5.02 |
| **931.29** | 44.68 ± 5.32 | 41.4 ± 6.04 | 33.85 ± 5.15 | 28.39 ± 4.44 | 22.19 ± 2.96 |
| **935.53** | 18.36 ± 2.50 | 23.32 ± 3.23 | 18.95 ± 2.33 | 16.56 ± 2.17 | 14.57 ± 1.85 |
| **1098.1** | 23.71 ± 3.39 | 17.52 ± 2.61 | 19.16 ± 2.48 | 20.64 ± 3.04 | 15.78 ± 2.34 |
| **1121.7** | 11.66 ± 1.69 | 12.98 ± 1.69 | 11.61 ± 1.61 | 10.47 ± 1.56 | 10.42 ± 1.80 |
| **1129.9** | 38.55 ± 5.66 | 34.22 ± 4.50 | 27.38 ± 3.66 | 22.08 ± 2.76 | 22.22 ± 3.29 |
| **1222.5** | 14.69 ± 2.20 | 13.28 ± 1.91 | 11 ± 1.70 | 9.46 ± 1.35 | 8.48 ± 1.31 |
| **1238.27** | 20.39 ± 2.92 | 19.08 ± 2.50 | 15.51 ± 2.63 | 11.89 ± 1.87 | 10.21 ± 1.90 |
| **1253.7** | 16.65 ± 2.33 | 13.11 ± 2.17 | 11.63 ± 1.85 | 9.62 ± 1.53 | 7.51 ± 1.40 |
| **1282.5** | 12.40 ± 1.75 | 9.55 ± 1.53 | 9.93 ± 1.51 | 12.26 ± 1.82 | 9.98 ± 1.70 |
| **1289.83** | 11.11 ± 1.70 | 11.78 ± 1.70 | 10.21 ± 1.45 | 9.11 ± 1.30 | 8.14 ± 1.16 |
| **1303.4** | 3.48 ± 0.55 | 3.56 ± 0.54 | 2.53 ± 0.46 | 2.14 ± 0.41 | 1.93 ± 0.36 |
| **1316.04** | 38.27 ± 5.81 | 35.41 ± 4.57 | 32.49 ± 3.99 | 27.55 ± 3.68 | 21.00 ± 3.10 |
| **1328.2** | 3.04 ± 0.50 | 2.70 ± 0.47 | 2.30 ± 0.42 | 2.17 ± 0.38 | 1.96 ± 0.37 |
| **1333.65** | 11.51 ± 1.64 | 15.35 ± 2.06 | 13.13 ± 1.93 | 11.72 ± 1.63 | 10.46 ± 1.48 |
| **1369.7** | 7.05 ± 1.00 | 6.49 ± 0.95 | 5.47 ± 0.91 | 4.80 ± 0.62 | |
| **1408.4** | 87.04 ± 13.64 | 84.34 ± 11.96 | 72.29 ± 9.63 | 57.82 ± 7.62 | 45.56 ± 7.28 |
| **1434.07** | 54.21 ± 7.84 | 64.29 ± 8.85 | 60.36 ± 7.70 | 51.69 ± 7.13 | 42.71 ± 6.05 |
| **1441.2** | 47.79 ± 5.64 | 36.53 ± 4.61 | 30.17 ± 4.14 | 25.04 ± 3.94 | 18.2 ± 2.92 |
| **1480.3** | 4.01 ± 0.67 | 7.19 ± 1.10 | 9.28 ± 1.56 | 7.81 ± 1.21 | 5.89 ± 0.96 |
| **1810.76** | 8.81 ± 1.36 | 7.01 ± 1.13 | 6.57 ± 1.04 | 5.17 ± 0.83 | 4.85 ± 0.77 |
| **2273.2** | 1.79 ± 0.31 | 1.47 ± 0.28 | 1.15 ± 0.21 | 0.96 ± 0.16 | 0.77 ± 0.14 |

### 4.3.1 Comparison to previous experimental data

In comparing the γ-ray line production cross section experimental data from this work (represented by red full circles in Figs. (5 – 12)) to previous counterparts, we considered the following series of data sets also reported in these figures:

- the data set measured by Dyer et al. [20] at the Washington University tandem accelerator up to $E_p$ = 23 MeV (shown in black, down-oriented full triangles),
- the data sets taken respectively by Benhabiles et al. [17] (shown in green circles) and Belhout et al. [24, 25] (green full diamonds) at the Orsay-Paris-Saclay University tandem accelerator up to $E_p$ = 26 MeV,
- the data set taken by Lesko et al. [29] (yellow, up-oriented full triangles) at the Lawrence Berkeley Laboratory's Cyclotron accelerators over the proton energy range of $E_p$ = 8.9 - 50 MeV and
- our data set taken [18] at the SSC of iThemba LABS for $E_p$ = 30 - 66 MeV (blue full squares).

As can be seen in Figs. (5 - 12), the previous experimental data sets from Refs. (17, 20, 24, 25) are mainly concentrated within the low proton energy region, our previous data set [18] extends the data from previous works over the proton energy range of $E_p$ = 30 - 66 MeV, while our results from the current work update and further extend the previous experimental data over the proton energy range of $E_p$ = 66 - 125 MeV. All our contributions, bringing substantial experimental cross section data, are particularly important as they reduce the need for extrapolation on a non-experimental basis over the proton high-energy region, then allowing to better identify the nuclear reaction processes behind the γ-ray production in astrophysical sites.

Observing Figs. (5 - 12) reveals that the cross section experimental data from this work and Refs. [17, 18, 20, 24, 25] show to be mutually consistent. However, the cyclotron data set of Lesko et al. [29] appears to significantly deviate upward relative to the other previous data sets for most of the γ-ray lines involving this data set. One also notes that only our data sets from this work and Ref. [18] are present in Figs. (5 - 12) for the majority of the γ-ray lines, these two data sets being mutually consistent within the involved experimental uncertainties. In the case of other γ-ray lines, one observes that only the data set from the current work is reported.

Further observation of Figs. (5 - 12) shows that for the lines having earlier experimental data reported (other than our previous data set from Ref. [18]), located below and around the low proton energy compound resonance structure at $E_\gamma$ = 20 - 40 MeV, an overall consistency holds between our two experimental data sets and lower energy data from previous works [17, 20, 24, 25], while the Lesko et al. [29] data set invariably appears to deviate upward from the other data sets. Since the latter data set [29] was taken into account in the Murphy et al. semi-empirical compilation [31], it consequently affects both in magnitude and in energy dependency the trends of γ-ray line excitation functions predicted by this compilation.

For further details, we proceed below to comparing separately the various experimental data sets for each explored target.

### A. γ-rays in proton reactions with the $^{nat}$Mg target

Among the cross section experimental results from the $^{nat}$Mg target presented in Figs. (5, 6), those for the three lines at $E_\gamma$ = 563.2, 937.0 and 1016.9 keV are reported here for the first time, to our knowledge. As can be observed in these figures, the cross section data often exhibit a gradual, more or less marked reduction as the proton beam energy increases beyond the compound resonance structure over the range of $E_p \cong$ 20 - 30 MeV. One can also see that the current study significantly extends to higher proton beam energies the results from previous works [18, 20, 24, 25, 29] with maintaining a direct continuity between them.

As can be checked in Table 1, the $^{24}$Mg is the most abundant isotope within the $^{nat}$Mg target relative to the $^{25}$Mg and $^{26}$Mg isotopes. The production cross sections for the γ-ray line at $E_\gamma$ = 1808.68 keV from $^{26}$Mg (E2 transition from the $J_i^\pi = 2^+$, 1$^{rst}$ excited state of this isotope to its ground state ($J_f^\pi = 0^+$, see Table 2), show to be in excellent agreement with our experimental data sets from this work and Ref. [18]. Notice also that the previous data sets from Refs. [24, 25] for this line are not included in Figs. (5, 6) due to substantial contributions from the Aluminum backing [25].

We also report in these figures the γ-ray production cross sections from the de-excitation of excited states in $^{25}$Mg. In particular, the γ-ray line at $E_\gamma$ = 585.03 keV (E2 transition from the $1/2^+$, 1$^{rst}$ excited state to the $5/2^+$ground state), and two other lines at $E_\gamma$ = 389.71 and 974.74 keV resulting from the de-excitation of the 2$^{nd}$ excited state of this isotope (see Table 2). The analysis of the γ-ray line at 585.03 keV must account for the contribution of another transition at $E_\gamma$ = 583.04 keV, probably resulting from the de-excitation of the $1^+$, 1$^{rst}$ excited state of $^{22}$Na to its $3^+$ ground state. For the γ-ray transitions at $E_\gamma$ = 389.71 and 974.74 keV, our experimental results are in good agreement with our previous data set [18]. In case of the line $E_\gamma$ = 585.03 keV, our cross sections from the current work are consistent with the data sets from Refs. [24, 25], but differ from our previous data set [18], which is higher by ≈ 22% at $E_p$ = 66 MeV.

Two γ-ray lines emitted by $^{24}$Mg were also investigated: the line at $E_\gamma$ = 1368.63 keV resulting from the decay of the 1$^{rst}$ excited state to the ground state, and the line at $E_\gamma$ = 2754.02 keV (E2 transition from the $4^+$, 2$^{nd}$ excited state to the $2^+$, 1$^{rst}$ excited state). The former line can be affected by another nearby line at Eγ = 1368.7 keV attributed to the $^{22}$Na isotope (transition from the $2^+$, 2$^{nd}$ excited state to the $1^+$, 1$^{rst}$ excited state). Although the short lifetime of the emitting state leads to a strongly Doppler-shifted and broadened γ-ray line expected to be hardly separated from its neighbor line from $^{24}$Mg, this contribution has been taken into account in our analysis (see Fig. 6). Comparing our γ-ray line production cross section data for these two lines to previous data, one observes a smooth extension of the latter to higher proton energies, along with an excellent agreement with the data sets from Refs. [20, 24, 25]. In contrast, the cross section data reported by Lesko et al. [29] are significantly higher than our cross section values, which can be attributed to the inclusion by these authors [29] of other isotopes in the composition of their targets, then contributing to substantially enhance the cross sections at lower proton energies.

Two other γ-ray lines at $E_\gamma$ = 450.7 and 425.81 keV emitted respectively by the $^{23}$Mg and $^{24}$Al isotopes during their decay to their ground states were also analyzed. As shown in Fig. 6, our cross section data for these lines are in excellent agreement with our previous data set [18] and the previous data sets from Refs. [24, 25].

Other γ-ray lines attributed to the de-excitations of the 1rst excited states of the $^{21.23.24}$Na isotopes were identified in the γ-ray energy spectra recorded from the $^{nat}$Mg target, for which the corresponding production cross sections were determined. These are the lines at $E_\gamma$ = 331.91, 439.99, 472.7 and 563.19 keV keV, the latter line being reported here for the first time. Our cross sections from the current work for these lines show mostly to be in good agreement with our previous data set [18] and the previous data sets from Refs. [18, 24, 25], except for the line at $E_\gamma$ = 439.99 keV for which our previous cross section data set [18] is higher by ≈ 31% than our data from the current work at $E_p$ = 66 MeV.

The production cross sections for the γ-ray line at $E_\gamma$ = 937.0 keV from the de-excitation of $^{18}$F and the line at $E_\gamma$ = 1016.9 keV from the de-excitation of $^{23}$Ne (also coming from $^{22}$Na, see Table 2) have been determined for the first time in this work, to our knowledge.

Finally, the γ-ray line production cross sections for the line at $E_\gamma$= 350.73 keV, stemming from the decay of the 1rst excited state of $^{21}$Ne to its ground state, were determined and compared with our previous data from Ref. [18]. A difference is observed between our current cross section data and our previous data [18], amounting to 24.7 % at the proton energy of $E_p$ = 66 MeV.

**B. γ-rays in proton reactions with the $^{nat}$Si target**

The proton irradiations of the $^{nat}$Si target led us to the identification of 15 γ-ray lines. Their analysis indicates that they are predominantly produced in binary nuclear reactions involving the $^{27}$Si, $^{28}$Si, $^{29}$Si, and $^{30}$Si isotopes making up the natural Si target (see Table 3). The corresponding experimental production cross sections were determined and are compared in Figs. (7, 8) to previous counterparts from Refs. [17, 18, 20, 29]. As in case of the $^{nat}$Mg target, one observes a progressive reduction of the production cross sections as the proton beam energy increases beyond the low energy region of compound resonances.

The production cross sections of the γ-ray line at $E_\gamma$ = 1273.36 keV, due to the de-excitation of the 1rst excited state of $^{29}$Si, were determined. The previous data sets for this line, reported in Refs. [17, 18], appear to be consistently extended to higher proton energies thanks to the current work.

The most intense line at $E_\gamma$ = 1778.97 keV, emitted in the de-excitation of $^{28}$Si, was also analyzed and its experimental production cross sections were determined. Previous experimental cross section data sets for this line have been reported in Refs. [17, 18, 20, 29].

The agreement between our data sets from this work and Ref. [18] is excellent at $E_p$ = 66 MeV.

As can be seen, these results are consistent with the previous lower proton energy data sets from Refs. [17, 20], but are slightly higher than that reported by Lesko et al. [29].

Two main γ-ray lines at $E_\gamma$ = 780.8 and 957.3 keV emitted in the de-excitations of $^{27}$Si were analyzed (see Table 3). As shown in Fig. 8, our cross section data sets for these lines from this work and Ref. [18] are consistent with the Lesko et al. data [29].

Two γ-ray lines at $E_\gamma$ = 416.85 and 843.76 keV coming from the de-excitation of $^{26,27}$Al isotopes were observed and their production cross sections were determined, the second line being also fed by another transition at $E_\gamma$ = 843.92 keV from the $^{26}$Al isotope. A very good agreement is observed between our cross section data sets for these lines from this work and Ref. [18]. However, concerning the 843.76 keV line the previous cross section data set reported in Ref. [29] at proton energies above $E_p$ = 30 MeV appears to be higher and exhibiting a different energy dependence compared to the overall datasets.

The γ-ray production cross sections were determined for transitions from the de-excitation of the 1$^{rst}$ two excited states of $^{25}$Mg, produced in proton binary reactions with the $^{nat}$Si target. They concern the lines at Eγ = 389.71, 585.03 keV and, for the first time, the line at Eγ = 974.74 keV line (see Table 3). The 585.03 keV line overlaps with another line of very close energy located at Eγ = 583.04 keV emitted during the de-excitation of $^{22}$Na. A very good agreement and a smooth continuity at high proton energies are observed between our data sets for these lines from this work and Ref. [18].

A line at $E_\gamma$ = 1368.63 keV from the $^{24}$Mg isotope was also observed and analyzed. As previously mentioned in the case of the $^{nat}$Mg target, another line with very close energy of $E_\gamma$ = 1368.7 keV, emitted by the $^{22}$Na isotope, is likely to overlap with this line of $^{24}$Mg (see Table 3). Our cross section data for this doublet are consistent with our previous data set [18], although the latter values are higher by about 30% to 40% relative to the proton lower energy data of Lesko et al. [29].

The γ-ray production cross sections were determined here for the first time for a second line at $E_\gamma$ = 890.87 keV from the $^{22}$Na isotope, for which no previous cross section data has been reported in the literature, to our knowledge.

We also determined the production cross sections for γ-ray lines emitted in the deexcitation of the 1$^{rst}$ excited states of $^{23}$Mg, $^{21,23}$Na and $^{21}$Ne isotopes, for which only our previous cross section data set [18] has been reported. These are the lines at $E_\gamma$ = 450.7 keV from $^{23}$Mg, $E_\gamma$ = 439.99 keV from $^{23}$Na, $E_\gamma$ = 331.91 keV from $^{21}$Na, and $E_\gamma$ = 350.73 keV from $^{21}Ne$ (see Table 3). As can be seen in Fig. 8, an excellent agreement is observed between our two experimental cross section data sets for these lines from this work and Ref. [18].

### C. γ-rays in proton reactions with the $^{56}Fe$ target

The iron target used in our experiments was highly enriched, to over 99%, in $^{56}Fe$. Figs. (9 - 12) report the production cross sections for 30 γ-ray lines produced mainly in inelastic proton scattering. Our experimental data for the lines at $E_\gamma$ = 125.95, 156.27, 211.98, 274.8, 377.88, 411.9, 846.77, 931.29, 1129.29, 1222.5, 1238.27, 1303.4, 1316.04, 1408.4, 1434.07, 1441.2, 1810.76 and 2273.1 keV are compared in these figures to the previous experimental data sets reported in Refs. [18, 20, 24, 25, 29]. One observes that the experimental γ-ray production cross sections show a marked increase over the low proton energy region, reaching maxima approximately in the energy range of $E_p$ = 10 - 60 MeV. This behavior is characteristic of nuclear reactions dominated by the compound nucleus mechanism. Beyond this region, a gradual decrease in the γ-ray cross sections is observed with increasing the proton beam energy. This trend signs the transition to direct and pre-equilibrium nuclear reaction mechanisms, and successive openings of other competitive reaction pathways, such as the emission of nucleons and alpha particles, which reduces the probability of γ-ray line emission. The production cross sections for the other γ-ray lines at $E_\gamma$ = 226.4, 783.3, 935.53, 1098.1, 1121.7, 1251.1, 1282.5, 1289.59, 1328.2, 1333.65, 1369.7 and 1480.3 keV (see Table 4) are reported for the 1rst time in this work, to our knowledge. As can be observed, our results show a clear continuity of the γ-ray production cross sections measured in this work with our previous data set [18] and the earlier data sets reported in Refs. [20, 24, 25, 29].

We determined the γ-ray production cross sections for the 1rst three levels of the $^{56}Fe$ mass spectral band. They correspond to the intense lines at 846.77, 1238.27 and 1810.76 keV emitted in the de-excitation of the 1rst three excited states of $^{56}Fe$. Our cross section results for these lines show good agreement with our previous data set [18] and with the lower-energy data sets of Belhout et al. [24, 25] and Dyer et al. [20]. In contrast, the data set of Lesko et al. [29] appears to be systematically higher than the previous data sets for the lines at $E_\gamma$ = 846.77 and 1238.27 keV at proton beam energies above $E_p$ = 30 MeV.

The extraction of the integral cross section for the 846.78 keV γ-ray line was carried out through a careful and rigorous deconvolution procedure aiming at removing the neutron induced background and interfering adjacent line peaks. A multipeak Gaussian fitting procedure was indeed employed to separate the contributions from the line peaks at $E_\gamma$ = 843.76 keV from $^{27}Al$ and $E_\gamma$ = 848 keV from $^{52}Cr$. The interfering Aluminum component arises from the inelastic scattering of secondary neutrons and protons off the AFRODITE reaction chamber constituent material. Similarly, in the fitting of the γ-ray line at 1238.27 keV, the contribution of a small spurious line peak at $E_\gamma$ = 1241.7 keV, attributed to the decay of $^{53}$Mn, was taken into account.

Besides, the cross sections for the γ-ray lines at $E_\gamma$ = 1303.4 and 2273.2 keV, also coming from the decay of this isotope (de-excitations of the 10$^{th}$ and 7$^{th}$ excited states, respectively), were determined. The obtained results are consistent both with our previous data set [18] and the earlier data sets from Refs. [24, 25].

Gamma-ray lines originating from the decays of the $^{53.54.55}$Fe isotopes were also observed in the γ-ray energy spectra from the $^{56}$Fe target, and the corresponding production cross sections were determined. These are the lines at $E_\gamma$ = 274.8, 1222.5, 1129.9 keV and, for the 1$^{rst}$ time, the lines at 1328.2 and 1369.7 keV (see Table 4).

Furthermore, production cross sections were determined for two γ-ray line doublets at $E_\gamma$ = 411.9 and 1408.4 keV attributed, respectively, to the de-excitation of the $^{54}$Fe and $^{55}$Fe isotopes.

The 1316.4 keV γ-ray line shows a low energy bump associated with a 1312.2 keV line, probably originating from $^{56}$Fe. Because of Doppler broadening, these two components could not be resolved. Then, their peak areas were combined (see Ref. [18]). The 931.29 keV γ-ray line is a triplet that may result from the de-excitation of the 5/2$^-$ excited state of $^{55}$Fe to its 3/2$^-$ state, in addition to the transitions: $5^+ \rightarrow 7^+$ and $8^+ \rightarrow 7^-$ in $^{54}$Mn.

It is always found that the γ-ray line production cross sections derived in this work show to be in good agreement with the previous experimental data reported in the literature, excepting the data set from Ref. [29]. In particular, this data set is distinguished by a significant discrepancy with other data for the γ-ray lines at $E_\gamma$ = 931.29, 1129.9, 1222.5 and 1408.4 keV, a difference becoming pronounced for incident proton beam energies above $E_p$ = 30 MeV.

The γ-ray production cross sections for the lines at $E_\gamma$ = 156.29, 211.98, 377.88 and 1441.2 keV, resulting from the de-excitation of the $^{53.\,54}$Mn isotopes, and the 125.95 keV line coming from the de-excitation of $^{55}$Mn were determined and compared to our data from this work and from Ref. [18] and to the previous ones from Refs. [24, 25]. Very good agreements along with an excellent continuity with these data sets is observed.

The production cross sections for several new γ-ray lines have been determined in this work. This includes lines originating from the de-excitation of the $^{52.53}$Mn isotopes, particularly the lines at Eγ = 1253.7, 1121.7 and 1289.83 keV, as well as six new lines at Eγ = 783.3, 935.53, 1098.1, 1282.5, 1333.65 and 1480.3 keV from the decay of the $^{50.51.52}$Cr isotopes. In addition, the 226.3 keV line from $^{50}$V has been determined. These new γ-ray line production cross sections, reported in Table 4, then constitute a decisive experimental contribution to the existing database [31] and allow a better understanding of the γ-ray emission patterns.

Finally, in case of the γ-ray line at $E_\gamma$ = 1434.07 keV emitted in the de-excitation of the 1rst excited state of $^{52}Cr$, a good agreement is observed between our cross section experimental data from the present work, our previous data set [18] and the previous data sets reported in Refs. [24, 25, 29].

### 4.3.2 Comparison to the Murphy et al. semi-empirical compilation

In 2009, Murphy et al. [31] compiled the available experimental cross section data for about 140 main, intense γ-ray lines prominent in the SFs and ISM, resulting from the decay of low-lying excited states produced in nuclear reactions induced by light ions (protons, $^3He^+$ ions and α-particles) on target nuclei ranging from $^4He$ to $^{56}Fe$. In particular, updating earlier databases for proton and α-particle induced reactions, they extrapolated the existing experimental data to higher projectile energies up to several hundreds of MeV/u not covered by experiment, based on TALYS code [32] calculations. In addition, they provided predicted trends of γ-ray line excitation functions for weak lines produced during the decay of high-lying excited states, also based on TALYS code calculations.

The cross section excitation functions predicted by the Murphy et al. semi-empirical compilation [31] are also plotted in Figs. (5-12) for the sake of comparison to the integral σ ($E_p$) experimental data from the present and previous works [17, 18, 24, 25, 29]. Observing these figures, one first notes that the predicted cross section curves show an overall consistency with the experimental data sets regarding their proton energy dependence for most of the γ-ray lines, while they show significant deviations in magnitude from them in the presence of the Lesko et al. data set [29]. One also observes that, when it is present, the experimental cross section data set of Lesko et al. [29] (considered in the Murphy et al. compilation) is often significantly higher than the other experimental data for $E_p \geq 30$ MeV probably due to the inclusion by these authors of more isotopes in their investigated targets. One may thus expect that in case of the γ-ray lines having earlier experimental cross section data sets included in the Murphy et al. compilation [31] up to $E_p \cong 50$ MeV (the highest proton energy in the Lesko et al. [29] data set), the predicted cross section curves should agree with the σ ($E_p$) experimental data in presence of the Lesko et al. data set [29]. However, this turns out to be true only in the case of two γ-ray lines at Eγ = 1368.63 + 1388.7 keV (doublet) and Eγ = 1778.97 keV produced in (p, p'γ) inelastic proton scattering off the $^{nat}Si$ target.

Otherwise, variable observations hold corresponding to the facts that the predicted cross section curves from Ref. [31] either overestimate or underestimate the measured cross section data in the presence or the absence of the Lesko et al. data set.

As can be seen in Figs. (5 - 12), the cross section curves predicted by the Murphy et al. compilation [31] show to be in partial agreement with the experimental data in case of the lines at Eγ = 1808.68 keV (in absence of the Lesko et al. data set [29]) from the $^{nat}$Mg target, at Eγ = 1368.63 + 1368.7 keV (in presence of the Lesko et al. data set [29]), at Eγ = 1778.97 keV (in presence of the Lesko et al. data set [29]) from the $^{nat}$Si target, and at Eγ = 1369.7 keV (in absence of the Lesko et al. data set [29]) from the $^{56}$Fe target.

The predicted curves overestimate the experimental cross section data in the presence of the Lesko et al. data set [29] for the lines at Eγ = 1388.63 + 1388.7 from the $^{nat}$Mg target, and Eγ = 846.77 + 847 keV, Eγ = 931.29 + 931.2 + 931.04 keV (triplet), Eγ = 1129.29 keV, Eγ = 1238.27 keV, Eγ = 1312.26 + 1318.4 keV keV, Eγ = 1408.4 + 1408.1 keV from the $^{56}$Fe target.

They also overestimate the cross section experimental data sets even in the absence of the Lesko et al. [29] data set in case of the two lines at Eγ = 1303.4 keV, Eγ = 1810.78 keV from the $^{56}$Fe target.

In contrast, the predicted curves by the Murphy et al. compilation [31] underestimate the cross section experimental data sets in the absence of the Lesko et al. [29] data set in case of the γ-ray lines at Eγ = 439.99 keV, Eγ = 450.7 keV and Eγ = 2754.12 keV from the $^{nat}$Mg target, and Eγ = 411.4 + 411.9 keV from the $^{56}$Fe target. They also underestimate the cross section experimental data sets in the presence of the data set of the Lesko et al. [29] in case of the γ-ray lines at Eγ = 780.8 keV, Eγ = 843.76 + 843.92 keV, Eγ = 957.3 keV from the $^{nat}$Si target, and Eγ = 1222.5 keV, Eγ = 1434.7 keV from the $^{56}$Fe target.

## 5. Interpretation within the nuclear reaction theory, TALYS code calculations

For interpreting and validating the experimental γ-ray line production cross sections, we compare them in this section to counterparts derived by TALYS code [32] calculations in terms of the predictions of nuclear reaction theory versus the incident energy domains covered by the accelerated protons.

This code allows calculating theoretical cross sections for nuclear reactions induced by γ-rays, neutrons and charged particles (p, d, t, $^3He^+$, α-particles) on various target nuclei over the incident energy range of $E_{lab}$ = 1 keV - 250 MeV, based on reaction mechanisms (compound nucleus, direct, pre-equilibrium, fusion-evaporation reactions, fission, etc.)

expected to occur over this incident energy region. It has been extensively used earlier by several groups for calculating nuclear cross sections for prompt and delayed γ-ray lines prominent in SFs and the ISM emitted in proton and α-particle induced reactions on stable and radioactive nuclei (see Refs.[18, 36] and references therein). It contains intrinsic nuclear data libraries (for masses, level densities, discrete states, OMP and $\beta_\lambda$ level deformation parameters, γ-ray branching ratios, strength functions, etc.) and default built-in parameters based on phenomenological or microscopic models. Alternatively, the user is allowed to introduce one's own input data (such as OMPs, nuclear level deformation parameters, etc.) referred to as modified TALYS code parameters that can be derived from theoretical models or the analysis of experimental data. This can be done using keywords provided in TALYS or the code's intrinsic nuclear data libraries.

Integrating several nuclear reaction models, this code proves to be a powerful tool for nuclear data prediction and analysis [16]. Enabling comparisons between its predictions and experimental data with allowing model adjustments, it provides predictions of nuclear observables over energy intervals that have not been explored experimentally. Several research groups [16, 17, 18, 30, 36, 38] have checked its efficiency and accuracy via comparing its predictions with experimental data for a variety of nuclear reactions.

At very low energies, nuclear reactions generally involve neutrons as projectiles due to the repulsive Coulomb barrier for charged particles, the dominant reaction mechanisms being elastic scattering and capture processes. As the incident energy increases and inelastic scattering channels open, the reactions proceed through compound nucleus formation and direct reactions. The latter are described by the Distorted Waves Born Approximation (DWBA) for spherical nuclei and the model of coupled-channel reactions in case of deformed nuclei. At higher energies, pre-equilibrium mechanisms and multiple particle emission become predominant. In this context, TALYS ensures a gradual transition between these different reaction mechanisms, thus enabling reliable modeling over a wide energy range where one or more mechanisms may simultaneously contribute depending on the considered physical conditions.

For the γ-ray emission, TALYS distinguishes two types of nuclear levels: (i) the $1^{rst}$ type corresponding to low-excitation, discrete nuclear levels that de-excite exclusively by γ-ray emission taken from the RIPL library [51], and (ii) a $2^{nd}$ type of higher-excitation levels, treated statistically using energy bins that group a large number of states. The latter levels are handled through a continuous level density rather than considered as individual states in order to account for their high density and collective behavior.

There exist several versions of the TALYS code, which is still under development. In the current work, we used the 1.96 version [32].

In a first step, we calculated theoretical γ-ray line cross section excitation functions using the default parameters of TALYS, with introducing only the incident energy values and the mass and atomic numbers of the projectiles and targets. The obtained results are reported in Figs (5 - 12) together with the measured cross section data.

Observing Figs (5 – 12), one first notes a gradual decrease in the calculated γ-ray production cross sections as the incident energy increases beyond the low energy region dominated by compound nucleus resonances, due to the successive openings of other competitive nuclear reaction channels. For the majority of the γ-ray lines, this trend is well followed by the experimental data. As can be seen, at some exceptions the calculated excitation functions account reasonably well for the energy dependence of the experimental data for most of the γ-ray lines. In some cases, they also account partially even quantitatively for the measured cross section data. This holds for the lines at Eγ = 331.91, 439.99, 472.2, 937.0, 1016.8 + 1016.9, 1368.63 + 1368.7 keV from the $^{nat}$Mg target, Eγ = 416.85, 450.7 + 451.7 keV from the $^{nat}$Si target, and Eγ = 125.95, 156.27, 211.98, 274.8, 847.77 + 847, 931.29 + 931.2 + 931.04, 1238.27, 1312.6 + 1318.4, 1441.2, 1480.3 keV from the $^{56}$Fe target. However, the calculated excitation functions show substantial discrepancies relative to the experimental data for several γ-ray lines.

Indeed, while at very low energies satisfactory agreements are observed between the calculated cross sections and experimental data, significant differences appear elsewhere over a wide energy range. This led us to modify the default input parameters in TALYS so that the calculated cross sections reproduce as well as possible the experimental data. For doing this, we treated each nucleus separately by implementing different options in the code associated with each involved reaction mechanism, consistently with the nature of the target nucleus.

Particular attention was paid to the γ-ray lines resulting from the de-excitation of the 1$^{rst}$ excited states of certain nuclei. These transitions are important because they are often well defined and allow for reliable comparisons between different experiments. This is the case, e.g., for the γ-ray lines at $E_\gamma$ = 350.73, 1368.63, and 1808.68 keV from the $^{nat}$Mg target, the lines at $E_\gamma$ = 843.76, 1368.63, and 1778.97 keV from the $^{nat}$Si target, and the lines at $E_\gamma$ = 846.77, 1238.27, 1316.04, 1408.4, and 1434.07 keV from the $^{56}$Fe target.

In the TALYS code, the contribution of the compound nucleus mechanism is mainly described by the Hauser-Feshbach's statistical model including width fluctuation corrections according to the Moldauer approach [52]. Direct reactions are treated within the framework of the DWBA approach through the ECIS-06 code [53]. Concerning the pre-equilibrium reactions, TALYS mainly uses the two-component exciton model developed by C. Kalbach [54]. The fitting of the parameters associated with these different reaction mechanisms shows that no significant differences is observed between the TALYS code predictions and the experimental data. These mechanisms are correctly formulated and implemented in TALYS. The modifications that produce a remarkable improvement and allow a better reduction of the discrepancy between the TALYS predictions and the cross section experimental data mainly concern the OMP and the $\beta_\lambda$ level deformation parameters. This is due to the fact that the TALYS code considers the nucleus as deformed, and that the calculations of the total γ-ray line production cross sections are performed using the deformed OMPs.

Then, in a second step, we performed TALYS code [32] calculations of the total γ-ray line cross sections with introducing our own modified OMP and $\beta_\lambda$ level deformation parameters in the code source, aiming to improve the agreements between the predicted theoretical excitation functions and the experimental data. This was done following the method detailed in our previous paper [18] reporting similar cross section results for the same targets of $^{nat}$Mg, $^{nat}$Si, and $^{56}$Fe explored over the proton beam energy range of $E_p$ = 30 - 66 MeV. For extracting these parameters, we have in priority analyzed tens of angular distribution experimental data for elastic and inelastic nucleon scatterings off the $^{24,25,26}$Mg, $^{28,29,30}$Si and $^{54,56}$Fe isotopes taken from the literature, e.g., from the IAEA's EXFOR library [55]. Then, we used the coupled-channels reactions code OPTMAN [56] available in the RIPL-3 data library [57] to generate comprehensive theoretical fits to the angular distribution experimental data for deriving our own modified OMP and $\beta_\lambda$ level deformation parameters. The form of the energy dependent interaction potential is given in Ref. [56]. In addition to optimizing the OMPs, this required us adjusting the nuclear level density parameters and the coupling schemes of γ-ray emitting residual isotopes listed in tables (2, 3, 4).

Another important improvement made concerns the inclusion in the calculations of many γ-ray line production cross sections, and of contributions from multiple particle emission such as the (p, pγ $^3$He) and (p, 2n$^3$He) reactions. These reactions produce the same γ-ray lines as the (p, p′) inelastic scattering and play a significant role, particularly at high incident energies. Their contributions have been neglected in the previous calculations. Tables 11 and 12 report, respectively, the optimized modified OMP and $\beta_\lambda$ level deformation parameters used in our TALYS code calculations.

The calculated γ-ray production excitation functions using our OMP and $\beta_\lambda$ level deformation parameters are also plotted (by solid pink curves) in Figs. (5 – 12) for comparison to the cross section experimental data from the present and previous works [17, 18, 20, 24, 25, 29]. As can be seen, they generally stand close to the cross section curves generated in our calculations using the default parameters of TALYS. Not surprisingly, the two calculation modes lead to mutually consistent trends of the cross section curves regarding their evolving proton energy dependances. However, more or less marked departures between the two sets of calculated cross section values are observed for several γ-ray lines with increasing the proton beam energy beyond the resonant structures at $E_p$ = 20 - 40 MeV.

Finally, one observes that significant improvements of the agreement between the calculated excitation functions using our modified OMP and $\beta_\lambda$ level deformation parameters and our experimental data (including our previous data set [18]) are obtained for most of the γ-ray lines. Good to excellent consistency is reached concerning the lines at $E_\gamma$ = 331.91, 350.73, 472.2, 583.04 + 583.03, 937.0+ 583. 03 , 1808.68 keV from the $^{nat}$Mg target, $E_\gamma$ = 389.71, 416.85, 450.7 + 451.7, 583.04 + 585.03, 843.76+ 843.92, 974.74, 1388.63 + 1388.7, 1778.87 from the $^{nat}$Si target, and $E_\gamma$ = 125.95, 156.27, 211.98, 274.8, 377. 88, 411.4 + 411.9, 846.77 + 847, 931.29 + 931.2 + 931.04, 935.53 + 936.9, 1222.5, 1238.27, 1251.1 + 1252.1 + 1253.7, 1289.59 + 1289.83, 1312.6 +1316.4, 1328.2 + 1328.95, 1333.05, 1369.7, 1408.4 + 1408.1, 1434.07, 1441.2, 1420.3, 2273.1, 2273.2 from the $^{56}$Fe target.

Below, we further discuss the obtained cross section results with considering separately the γ-ray lines produced in proton irradiation of each target system.

### A. γ-ray lines in proton reactions with the $^{nat}$Mg target

In Figs. (5 - 6), the solid pink curves represent the sum of the contributions from all isotopes composing the $^{nat}$Mg target. These figures clearly show that the use of our modified OMP and $\beta_\lambda$ level deformation parameters considerably improve the reproduction of the experimental data both in absolute value and in energy dependence over the entire investigated proton energy range. Especially for the lines at $E_\gamma$ =1808.68 keV (of $^{26}$Mg), 585.03 and 974.74 keV ($^{25}$Mg), 472.2 keV ($^{24}$Na), 331.91 keV ($^{21}$Na), 1016.9 keV ($^{23}$Ne), 350.73 keV ($^{21}$Ne) and 937.0 keV ($^{18}$F). Furthermore, for the six lines at $E_\gamma$ = 389.71 keV ($^{25}$Mg), 1368.63 and 2754.02 keV ($^{24}$Mg), 450.7 keV ($^{23}$Mg), 425.81 ($^{24}$Al) and 439.99 keV ($^{23}$Na), the agreement between the calculated cross section values and experimental data is in general quite good. A certain competition is observed between the calculations performed using the default TALYS parameters and those made with our modified parameters. For the lines at $E_\gamma$ = 389.71, 1368.63 and 425.81 keV, the calculated values with our modified parameters are higher than those generated with the TALYS default parameters, while the opposite situation is observed for the lines at $E_\gamma$ = 450.7 and 439.99 keV. Concerning the line at $E_\gamma$ = 2754.02 keV, the calculated values with the default parameters are higher for incident energies of $E_p$ < 60 MeV, while those calculated using our own modified parameters are higher over the remaining proton energy range. It is important to highlight the significant contribution of the $^{24}$Mg (p, x) $^{22}$Na reaction in the evaluation of the final cross sections for the γ-ray lines at $E_\gamma$ = 583.04, 1368.7, and 1016.8 keV.

However, both the excitation curves generated using the default parameters of TALYS and our own modified parameters are unable to reproduce the experimental data for the line at $E_\gamma$ = 563.19 keV ($^{24}$Na), that shows significant discrepancies between the calculated and measured cross section values.

Table 11. Adjusted values of our OMP parameters used as modified inputs in our TALYS calculations for proton induced reactions with $^{24,25,26}$Mg, $^{28,29,30}$Si and $^{56}$Fe.

| OMP Parameters | $^{56}$Fe | $^{28,30}$Si | $^{29}$Si | $^{24,26}$Mg | $^{25}$Mg |
|---|---|---|---|---|---|
| $E_f$ (MeV) | -6.96 | -7,16 | -7.16 | -7.16 | -6.18 |
| $r_c$ (fm) | 1,361 | 1,324 | 1.324 | 1.324 | 1.324 |
| $r_v$ (fm) | 1.282 | 1.190 | 1.160 | 1.184 | 1.196 |
| $a_v$ (fm) | 0.665 | 0.552 | 0.669 | 0.562 | 0.592 |
| $V_1$ (MeV) | 60.2 | 51.2 | 51.2 | 50.3 | 60.5 |
| $V_2$ (MeV $^{-1}$ ) | 0.0082 | 0.0051 | 0.0055 | 0.0053 | 0.0055 |
| $V_3$ (MeV $^{-2}$ ) | 0.000033 | 0.000022 | 0.000018 | 0.000022 | 0.000017 |
| $W_1$ (MeV) | 14.2 | 14.0 | 15.0 | 14.0 | 14.0 |
| $W_2$ (MeV) | 75.0 | 75.0 | 75.0 | 75.0 | 75.0 |
| $r_{vd}$ (fm) | 1.290 | 1.190 | 1.160 | 1.184 | 1.196 |
| $a_{vd}$ (fm) | 0.585 | 0.522 | 0.590 | 0.562 | 0.592 |
| $d_1$ (MeV) | 15.0 | 14.6 | 12.5 | 14.6 | 14.6 |
| $d_2$ (MeV $^{-1}$ ) | 0.0580 | 0.0216 | 0.0216 | 0.0216 | 0.0216 |
| $d_3$ (MeV) | 10.00 | 11.10 | 11.10 | 11.10 | 11.10 |
| $r_{vso}$ (fm) | 1.000 | 1.123 | 1.007 | 1.078 | 1.078 |
| $a_{vso}$ (fm) | 0.580 | 0.757 | 0.746 | 0.815 | 0.580 |
| $V_{so1}$ (MeV) | 8.1 | 7.7 | 7.9 | 7.7 | 9.7 |
| $V_{so2}$ (MeV $^{-1}$ ) | 0.0090 | 0.0035 | 0.0035 | 0.0035 | 0.0035 |
| $W_{so1}$ (MeV) | -3.1 | -4.5 | -4.5 | -4.2 | -3.1 |
| $W_{so2}$ (MeV) | 160.0 | 332.0 | 160.0 | 223.0 | 160.0 |

Table 12. Adjusted values of the deformation parameters $\beta_\lambda$ for proton reactions with $^{24,25,26}$Mg, $^{28,29,30}$Si and $^{56}$Fe used as modified input in the TALYS calculations.

| **$\beta_\lambda$ deformation parameters** | **$^{24}$Mg** | **$^{25}$Mg** | **$^{26}$Mg** | **$^{28}$Si** | **$^{29}$Si** | **$^{30}$Si** | **$^{56}$Fe** |
|---|---|---|---|---|---|---|---|
| $\beta_2$ | 0.758 | 0.554 | 0.449 | -0.419 | -0.320 | 0.315 | 0.141 |
| $\beta_4$ | -0.037 | 0.152 | 0.182 | 0.108 | 0.250 | 0.220 | 0.001 |

### B. γ-ray lines in proton reactions with the $^{nat}$Si target

As in case of the $^{nat}$Mg target, the continuous pink curves shown in Figures 7 - 8 represent the sum of the contributions from the $^{28}$Si, $^{29}$Si, and $^{30}$Si isotopes constituting the natural $^{nat}$Si target. All the γ-ray lines from this target correspond to transitions from the 1rst excited state to the ground state, except for the 389.71 keV line of $^{25}$Mg, which corresponds to a transition from the 2nd to the 1rst excited state.

The excitation functions for the lines from the $^{nat}$Si target at $E_\gamma$ = 1778.97 ($^{28}$Si), 843.76 ($^{27}$Al), 416.85 ($^{26}$Al), 389.71 and 585.03($^{25}$Mg), 1368.63 keV ($^{24}$Mg), calculated using our modified OMP and $\beta_\lambda$ level deformation parameters, show an excellent agreement with the entire set of experimental data, both in the absolute values and the energy dependence over the whole investigated proton energy range. A similar observation also holds for the line at $E_\gamma$ = 450.7 keV (from $^{23}$Mg), especially when the contribution of the γ-ray line at $E_\gamma$ = 451.7 keV, originating from the de-excitation of the 1rst excited state of $^{25}$Al, is also taken into account, which had not been considered in previous studies.

In the case of the two new lines at $E_\gamma$ = 974.74 keV ($^{25}$Mg) and 890.87 keV ($^{22}$Na), the excitation functions calculated using our modified parameters are in very good agreement with the experimental data, fitting them very satisfactorily. For the lines located at $E_\gamma$ = 439.99 (from $^{23}$Na), 331.91 (from $^{21}$Na), and 350.73 keV ($^{21}$Ne), the calculated excitation functions using our modified input parameters show good overall agreement with the experimental data, correctly reproducing their absolute values for proton energies above 80 MeV. Conversely, for the lines of $^{27}$Si at $E_\gamma$ = 780.8 and 957.3 keV, the calculated excitation functions using our modified parameters show reasonable agreement with previous experimental data at proton energies below 45 MeV [18,29], but deviate significantly from the measured data for higher energies.

The only exception to the good results obtained in our study of the γ-ray lines from this target concerns the line of $^{29}$Si at $E_\gamma$ = 1273.36 keV; for this line, the calculations performed using our modified parameters and with the default parameters of TALYS show significant discrepancies with respect to the experimental data.

### C. γ-ray lines in proton reactions with the $^{56}$Fe target

Figs. 9 -12 show that the performed TALYS code calculations using our modified OMP and the $\beta_\lambda$ level deformation parameters provide theoretical excitation functions that describe very well the experimental data sets for most of the reported γ-ray lines produced in proton induced reactions on the $^{56}$Fe target.

For the 15 lines located at $E_\gamma$ = 846.77, 1238.27, 2273.02 ($^{56}$Fe), 274.8, 411.9, 931.29, 1222.5, 1316.04, 1408.4 ($^{55}$Fe), 377.88, 1441.2 ($^{53}$Mn), 156.29, 211.98 ($^{54}$Mn), 125.95 ($^{55}$Mn) and 1434.07 ($^{52}$Cr) keV, an excellent agreement is observed between the theoretical cross section values and the experimental data over the covered entire incident proton energy range.

The calculated γ-ray production cross sections for these transitions reproduce very well both our experimental data sets and those reported in Refs. [18, 20, 24, 25]. However, differences are sometimes observed between these experimental data sets and those reported in Ref. [29].

For the 2 γ-ray lines located at $E_\gamma$=1303.4 (from $^{56}$Fe) and 1129.9 ($^{54}$Fe) keV, the theoretical values show a good agreement, in general, with the experimental data. However, for the line at $E_\gamma$ = 1303.4 keV, the calculated values show to be slightly higher than the experimental values from this work and those reported in Ref. [18]. A similar observation also holds for the line at $E_\gamma$ = 1129.9 keV; however, in this case, our TALYS code calculations reproduce very well the experimental data reported in Refs. [24, 25].

For the 7 new lines at energies $E_\gamma$ = 935.53, 1333.65 ($^{52}$Cr), 1253.7 ($^{52}$Mn), 1289.83 ($^{53}$Mn), 1328.2 ($^{53}$Fe), 1369.7 ($^{56}$Fe) and 1480.3 ($^{51}$Cr) keV, an excellent agreement is observed between the calculated excitation functions using our modified input parameters and our experimental values, both in absolute values and in energy dependences. But for the 4 new lines at energies $E_\gamma$ = 1121.7 ($^{53}$Mn), 783.3, 1098.1 and 1282.5 ($^{50}$Cr) keV, the calculated cross sections using our modified input parameters overestimate our experimental data without showing any significant improvement compared to the calculations performed with the default parameters of TALYS. It should be noted that the good results obtained for the 2 lines at $E_\gamma$ = 931.29 et 1253.7 keV are achieved when the contributions of 2 other distinct transitions of the same energies are taken into account, as indicated in Table 4. These contributions have not been considered in previous studies. It should be also noted that additional contributions for some lines are well predicted by our TALYS code calculations using our modified parameters: this is the case, e.g., of the lines at $E_\gamma$ = 2273.2, 411.9 and 1316.04 keV.

The cross sections for the line at $E_\gamma$ = 266.3 keV correspond to the de-excitation of the 1$^{rst}$ excited state of $^{50}$V. However, our TALYS code calculations for this line do not provide satisfactory agreement with experimental data. Besides, significant discrepancies are observed between the calculated values using our modified input parameters and the entire set of experimental data for the line at $E_\gamma$ = 1810.76 keV (from $^{56}$Fe), while the calculated cross sections using the default parameters of TALYS considerably underestimate the corresponding experimental data.

Finally, the predicted theoretical excitation functions for the $^{nat}$Mg, $^{nat}$Si and $^{56}$Fe explored in this work show an overall good agreement with the experimental data, despite some observed discrepancies that can be attributed to the influence of secondary nuclear reactions or to the presence of non-identified impurities within the explored targets.

These results significantly contribute to the understanding of the investigated nuclear reactions and to enrich the existing nuclear databases which are essential tools both in theoretical studies and for practical applications.

## 6. Discussion, summary, conclusions and perspectives

Ground laboratory nuclear γ-ray line production cross sections are required in various application purposes in nuclear physics, related interdisciplinary sciences, γ-ray astronomy, astrophysics and space sciences.

Our experimental γ-ray production cross section data sets from the current work and our previous experiments [18, 36 - 39], measured at the SSC facility of iThemba LABS, cover a wide proton energy range extending from $E_p$ = 30 up to 200 MeV. They were collected in proton irradiations of $^{nat}C$, Mylar ($^{12}C$ + $^{16}O$), $^{nat}Mg$, $^{nat}Si$, $^{40}Ca$ and $^{56}Fe$ solid targets of elements abundant in astrophysical sites, in particular SFs and the ISM for which main, prominent γ -ray lines provide crucial information on their structure and elemental composition.

In this work, we report total cross sections for 60 nuclear γ-ray lines produced in interactions of $E_p$ = 66 - 125 MeV proton beams with $^{nat}Mg$, $^{nat}Si$ and $^{56}Fe$ solid targets. The corresponding γ-ray energy spectra exhibit relatively narrow lines, less affected by Doppler effect in comparison to the energy spectra from the $^{nat}C$ and Mylar targets [36], strongly affected by Doppler-shifts and broadenings. E.g., we mention here in passing the broad, strong and prominent in SFs γ-ray lines of $^{12}C$ and $^{16}O$ at $E_\gamma$ = 4.439 and 6.129 MeV corresponding, respectively, to the E2 ($J^\pi = 2^+$, $E_x$ = 4.439 MeV, τ = 42 fs →$0^+$, g.s.) and E3 ($3^-$, 6.129 MeV, τ = 18.4 ps→$0^+$, g.s.) nuclear transitions. Produced mainly in (p, p') and (α, α') inelastic scatterings off the latter two elements and in reactions of accelerated $^{12}C$ and $^{16}O$ ions with abundant Hydrogen and Helium, they require special line-shape analyses (see Refs. [36, 50] and references therein), and allow to extract precious information on the properties of the accelerated particle populations, the underlying acceleration mechanisms and the magneto-hydrodynamic structure of the astrophysical sites [9 - 13]. An additional experimental data set, taken over the proton energy range of $E_p$ = 125 - 200 MeV, is currently under treatment.

The motivations of our studies rely on various applications and outcomes. In particular, with this large collection of nuclear γ-ray line production cross sections, the available nuclear databases, especially the Murphy et al. compilation [31], will be considerably enriched and up-dated. This would allow further applications, particularly in nuclear reaction theory, γ-ray astronomy and nuclear astrophysics for addressing diverse questions needing the knowledge of these observables and of the total γ-ray line emission fluxes [12, 13, 18, 26].

In Nuclear physics and related interdisciplinary sciences, the new experimental MeV nuclear γ-ray line production cross sections can allow to reliably improve several research topics. Primarily, modern theoretical nuclear reaction codes such as TALYS [32], EMPIRE2 [58] and OPTMAN [56] can be updated, notably for evaluating needed nuclear structure data and the OMPs, and for modeling nuclear interactions. They can be used for optimizing and benchmarking nuclear reaction models, particularly over the transition region between compound nucleus resonance and pre-equilibrium reactions [18, 36]. Besides, these data sets can be used in Ion Beam Analysis (IBA) for particle-induced Gamma-ray emission (PIGE) to conduct non-destructive depth profiling analyses in materials science and archeology, testing and/or developing experimental fusion reactors, and in medical physics. They can be also exploited in prompt γ-ray imaging (PGI) and directly used in Monte Carlo simulation codes, such as GEANT4 [45] for ensuring real-time and spatial monitoring to precisely determine where the therapeutic proton beam should stop inside the body of the treated patient.

Regarding the implications of MeV nuclear γ-ray line production cross sections in astrophysics and space sciences, many issues can be cited, including the following relevant topics:

(i) SFs and the ISM studies.

Mg, Si, and Fe, known as *alpha and iron-peak elements*, are highly abundant in these and other astrophysical sites. The new γ-ray production cross sections can be directly used for modeling the nuclear processes at work in these sites, which enables to study, e.g., their composition, the flare dynamics and the energy spectra of flare-accelerated protons [1-8, 12, 16, 18, 26, 31, 36, 50, 59, 60]. Besides, γ-ray diagnostics can be made using prominent lines (the 6129 keV line of $^{16}O$, 1369 keV line of $^{24}Mg$, 1779 keV line of $^{28}Si$, and 847 keV line of $^{56}Fe$) produced in cosmic-ray interactions with the ISM constituents. Performing γ-ray line shape analysis [36, 50, 59] provides crucial information on the properties of the ISM dust grains.

(ii) Interpretation of space missions' data.

The γ-ray experimental cross-section data are crucial for interpreting high-energy observations from past and future NASA's and ESA's programs [1-8, 60, 62], such as the SMM, INTEGRAL, FERMI, and e-ASTROGAM missions.

(iii) Cosmic ray studies.

Accurate γ-ray line cross sections can help modeling the interactions of CRs and LECRs with the interstellar dust and gas, providing insights into elemental abundances and cosmic-ray intensity in the galaxy [12, 60, 74].

(iv) Stellar nucleosynthesis.

These data support a better understanding of the production, destruction and spallation of intermediate-mass elements in astrophysical sites [63, 64, 73].

(v) Supernovae and radioactive decay chains.

Measurements of γ-ray lines from isotopes such as $^{56}$Ni, $^{56}$Co, and $^{56}$Fe can allow direct observation of the nucleosynthesis in supernovae and calculation of synthesized iron-peak element masses [60, 64 - 66].

(vi) Solar flare space weather diagnostics.

Acting as a particle accelerator, the Sun initiates nuclear reactions on abundant elements with production of γ-ray lines. These de-excitation lines together with the characteristic line at $E_\gamma$ = 2.223 MeV from neutron-capture on Hydrogen can be used to infer the properties of solar cosmic rays and the density of the photosphere [67 - 71].

(vii) ISM evolution and steady-state nucleosynthesis.

Long-lived radioactive isotopes, like the $^{26}$Al producing the line at $E_\gamma$ = 1.809 MeV ($T_{1/2}$ = 7 x $10^5$ years) and the $^{60}$Fe (lines at $E_\gamma$ = 1.173 and 1.333 MeV, $T_{1/2}$ = 2.62 × $10^6$ years) can serve as clocks for tracing chemical enrichment and ongoing nucleosynthesis within the galaxy [60; 72, 73].

(viii) Planetary geochemistry mapping.

The γ-ray spectrometers mounted onto planetary orbiters, like Mars Odyssey and Messenger, use the γ-ray line production cross sections for mapping the elemental composition of planetary surfaces [75, 76].

(ix) Classical novae and kilo-novae.

Precision γ-ray line production cross section data support the modeling of radioactive decay lines such as, e.g., the 511 keV annihilation flash from $^{18}$F and r-process elements in kilo-novae that can power the light curves of these explosive events [77, 78].

(x) Gamma-ray bursts (GRBs).

Several decades after their fortuitous discovery in the late 1960's, these most energetic and brightest γ-ray flashes in the universe originating from distant galaxies are still explored and debated, as pointed out recently in comprehensive studies [79, 80] of these extreme cosmic explosions. In particular, an elusive thermal component has been pointed out [79] in the GRB prompt emission spectrum observed by the GBM instrument on board the Fermi space telescope [3].

Laboratory MeV γ-ray line production cross sections are thus very useful. They can be pertinently and widely applied, especially to better addressing astrophysical and space science topics. Therefore, upon completing the analysis and validation of all our experimental cross section data sets, we could focus on some of them for investigating further related relevant outcomes with relying on up-to-date nuclear γ-ray line production cross sections of improved accuracy. In this perspective, comprehensive studies of the type of those dealt with in Refs. [4 - 16] on non-thermal accelerated particles like CRs and LECRs and their interactions with abundant constituents of astrophysical sites are very welcome, and deserve to be more confidently undertaken with considering the new, improved γ-ray line production cross section data. Hence, using both the current and previous [17, 18, 20, 21, 24, 25, 26, 28, 29] experimental γ-ray line production cross section data sets together with well suited LECRs energy spectra [81, 82] for calculating the total γ-ray emission flux, one can derive more reliable, up-to-date and significantly improved results to be compared and discussed relative to counterpart observational data from space telescope missions [1 - 8] and to the extended, high-energy emission flux from the decay in two γ-rays of the $\pi^0$ meson [83].

The experimental results presented in this work thus represent a significant advancement for the scientific community. They provide up-to-date and enriched nuclear databases, allow to improve the modeling of nuclear reactions, and to deepen fundamental questions in astrophysics and γ-ray astronomy. In this respect, our ongoing analysis of the experimental cross section data set taken over the highest proton energy range of $E_p$ = 125 - 200 MeV will further complete this work, opening new areas for innovative technologies in these scientific research fields, and helping to further interpret the γ-ray production and observations in the Universe. Finally, the precise knowledge of MeV γ-ray production cross sections appears as a cross-disciplinary tool at the intersection of fundamental nuclear physics and engineering, medicine, γ-ray astronomy and astrophysics. It will certainly continue to play a key role in understanding and exploiting nuclear phenomena both on Earth and in the Universe.

**References**

[1] The e-ASTROGAM Collaboration, De Angelis, A., Tatischeff, V. et al., Exploring the extreme Universe with gamma rays in the MeV – GeV range, The e-ASTROGAM mission. Exp Astron 44, 25–82 (2017). https:/ oi.org/10.1007/s10686-017-9533-6

[2] The e-ASTROGAM Collaboration, Vincent Tatischeff, Alessandro De Angelis et al, e-ASTROGAM mission: a major step forward for gamma-ray polarimetry, Journal of

Astronomical Telescopes, Instruments, and Systems, Vol. 4, Issue 1, 011003 (November 2017). https://doi.org/10.1117/1.JATIS.4.1.011003
[3] A. Abdo, M. Ackermann, A. Mea: Fermi observations of GRB 090902b: a distinct spectral component in the prompt and delayed emission. Astrophys. J. 706(1), L138 (2009) ; M. Ackermann, M. Ajello, A. Albert, A. Allafort, L. Baldini, G. Barbiellini, D. Bastieri, K. Bechtol, R. Bellazzini, E. Bissaldi, et al., High-energy gamma-ray emission from solar flares: summary of Fermi large area telescope detections and analysis of two M-class flares, Astrophys. J. 787(1) (2014) 15.
[4] R. J. Murphy, R. Ramaty, B. Kozlovsky, Solar Abundances from Gamma-Ray Spectroscopy: Comparisons with Energetic Particle, Photospheric, and Coronal Abundances, AIP Conference Proceedings, vol.232, American Institute of Physics, 1991, pp.439–444.
[5] G.H. Share, R.J. Murphy, Accelerated and ambient He abundances from gamma-ray line measurements of flares, Astrophys. J. 508(2) (1998) 876.
[6] R.C. Lin, S. Krucker, G.J. Hurford, D.M. Smith, H.S. Hudson, G.D. Holman, R.A. Schwartz, B.R. Dennis, G.H. Share, R.J. Murphy, et al., RHESSI observations of particle acceleration and energy release in an intense solar gamma-ray line flare, Astrophys. J. 595(2) (2003) L69.
[7] J. Kiener, M. Gros, V. Tatischeff, G. Weidenspointner, Properties of the energetic particle distributions during the October 28, 2003 solar flare from INTEGRAL/SPI observations, Astron. Astrophys. 445(2) (2006) 725–733.
[8] M.J. Harris, V. Tatischeff, J. Kiener, M. Gros, G. Weidenspointner, High resolution γ-ray spectroscopy of flares on the East and West limbs of the Sun, Astron. Astrophys. 461(2) (2007) 723–729.
[9] R. Ramaty, R.A. Schwartz, S. Enome, H. Nakajima, Gamma-ray and millimeter-wave emissions from the 1991 June X-class solar flares, Astrophys. J. 436 (1994) 941–949.
[10] R.J. Murphy, B. Kozlovsky, G.H. Share, X.-M. Hua, R.E. Lingenfelter, using gamma-ray and neutron emission to determine solar flare accelerated particle spectra and composition and the conditions within the flare magnetic loop, Astrophys. J. Suppl. Ser. 168(1) (2007) 167.
[11] A.S.L. Sironi, Relativistic reconnection: an efficient source of non-thermal particles, Astrophys. J. Lett. 783(1) (2014) L21.
[12] H. Benhabiles-Mezhoud, J. Kiener, V. Tatischeff and A.W. Strong, Deexcitation nuclear gamma-ray line emission from low-energy cosmic rays in the inner Galaxy, The Astrophysical Journal, 763:98 (8pp), 2013 February 1; Erratum: "Deexcitation nuclear gamma-ray line emission from low-energy cosmic rays in the inner Galaxy" (2013, ApJ, 763, 98), Astrophys. J. 766(2) (2013) 139.

[13] T. Vieu, S. Gabici, V. Tatischeff, Non-linear particle reacceleration by multiple shocks, Mon. Not. R. Astron. Soc. 510(2) (2022) 2529–2537.
[14] R. Ramaty, B. Kozlovsky, R.E. Lingenfelter, Nuclear gamma-rays from energetic particle interactions, Astrophys. J. Suppl. Ser. 40 (1979) 487–526.
[15] B. Kozlovsky, R. Murphy, R. Ramaty, Nuclear deexcitation gamma-ray lines from accelerated particle interactions, Astrophys. J. Suppl. Ser. 141(2) (2002) 523.
[16] R.J. Murphy, B. Kozlovsky, J. Kiener, G.H. Share, Nuclear gamma-ray de-excitation lines and continuum from accelerated-particle interactions in solar flares, Astrophys. J. Suppl. Ser. 183(1) (2009) 142.
[17] H. Benhabiles-Mezhoud, J. Kiener, J.-P. Thibaud, V. Tatischeff, I. Deloncle, A. Coc et al., Measurements of nuclear γ-ray line emission in interactions of protons and α-particles with N, O, Ne, and Si, Phys. Rev. C 83(2) (2011) 024603 ;
[18] W. Yahia-Cherif, S. Ouichaoui, J. Kiener, E.A. Lawrie, J.J. Lawrie, V. Tatischeff, A. Belhout, D. Moussa, P. Papka, H. Benhabiles-Mezhoud, et al., Measurement and analysis of nuclear γ-ray production cross sections in proton interactions with Mg, Si, and Fe nuclei abundant in astrophysical sites over the incident energy range E= 30–66 MeV, Phys. Rev. C 102(2) (2020) 025802.
[19] W.T. Vestrand, G.H. Share, R.J. Murphy, D.J. Forrest, E. Rieger, E.L. Chupp, G. Kanbach, The solar maximum mission atlas of gamma-ray flares, Astrophys. J. Suppl. Ser. 120 (2) (1999) 409.
[20] P. Dyer, D. Bodansky, A.G. Seamster, E.B. Norman, D.R. Maxson, Cross sections relevant to gamma-ray astronomy: proton induced reactions, Phys. Rev. C 23(5) (1981) 1865.
[21] J. Narayanaswamy, P. Dyer, S.R. Faber, S.M. Austin, Production of 6.13-MeV gamma rays from the $^{16}O$ (p, p γ) $^{16}O$ reaction at 23.7 and 44.6 MeV, Phys. Rev. C 24(6) (1981) 2727.
[22] A.G. Seamster, E.B. Norman, D.D. Leach, P. Dyer, D. Bodansky, Cross sections relevant to gamma-ray astronomy: alpha-particle-induced reactions, Phys. Rev. C 29(2) (1984) 394.
[23] P. Dyer, D. Bodansky, D.D. Leach, E.B. Norman, A.G. Seamster, Cross sections relevant to gamma-ray astronomy: alpha-particle-induced reactions on $^{12}C$, $^{14}N$, and $^{16}O$ nuclei, Phys. Rev. C 32(6) (1985) 1873.
[24] A. Belhout, J. Kiener, A. Coc, J. Duprat, C. Engrand, C. Fitoussi, M. Gounelle, A. Lefebvre-Schuhl, N. De Séréville, V. Tatischeff, et al., γ-ray production by proton and α-particle induced reactions on $^{12}C$, $^{16}O$, $^{24}Mg$, and Fe, Phys. Rev. C 76(3) (2007) 034607.
[25] A. Belhout, J. Kiener, A. Coc, J. Duprat, C. Engrand, C. Fitoussi, M. Gounelle, A. Lefebvre-Schuhl, N. De Séréville, V. Tatischeff, et al., Erratum: γ-ray production by proton

and α-particle induced reactions on $^{12}C$, $^{16}O$, $^{24}Mg$, and Fe [Phys. Rev. C 76, 034607 (2007)], Phys. Rev. C 80(2) (2009) 029902.
[26] H. Benhabiles-Mezhoud, Calcul du spectre total de l'émission gamma induite par interactions nucléaires des particules du rayonnement cosmique avec le milieu interstellaire et comparaison avec les observations de l'astronomie gamma, Ph.D. thesis, Paris 11, 2010.
[27] J. Kiener, A. Belhout, V. Tatischeff, H. Benhabiles-Mezhoud, in: Proceedings of the DAE Symposium on Nuclear Physics, vol.53, Bhabha Atomic Research Centre, Roorkee, India, 2008.
[28] J. Kiener, M. Berheide, N.L. Achouri, A. Boughrara, A. Coc, A. Lefebvre, F. de Oliveira Santos, C. Vieu, γ-ray production by inelastic proton scattering on $^{16}O$ and $^{12}C$, Phys. Rev. C 58(4) (1998) 2174.
[29] K.T. Lesko, E.B. Norman, R.-M. Larimer, S. Kuhn, D.M. Meekhof, S.G. Crane, H.G. Bussell, Measurements of cross sections relevant to -ray line astronomy, Phys. Rev. C 37(5) (1988) 1808.
[30] J. Kiener, J. Bundesmann, I. Deloncle et al., γ-ray emission in α-particle interactions with C, Mg, Si, and Fe at E = 50 - 90 MeV, Phys. Rev. C 104(2) (2021) 024621.
[31] R.J. Murphy, B. Kozlovsky, J. Kiener, G.H. Share, Nuclear gamma-ray de-excitation lines and continuum from accelerated-particle interactions in solar flares, Astrophys. J. Suppl. Ser. 183(1) (2009) 142.
[32] A.J. Koning, S. Hilaire, M.C. Duijvestijn, TALYS-1.0, in: International Conference on Nuclear Data for Science and Technology, EDP Sciences, 2007, pp.211–214 ; A.J. Koning, S. Hilaire, and S. Goriely. "TALYS: modeling of nuclear reactions." *The European Physical Journal A* 59.6 (2023): 1-85.
[33] J. Sharpey-Schafer, Laboratory portrait: iThemba laboratory for accelerator-based sciences, Nucl. Phys. News 14(1) (2004) 5–13.
[34] R. Newman, J. Lawrie, B. Babu, M.S. Fetea, S.V. Förtsch, S. Naguleswaran, J.V. Pilcher, D.A. Raave, C. Rigollet, J. Sharpey-Schafer, C.J. Stevens, F.D. Smit, G.F. Steyn, C.V. Wikner, D.G. Aschman, R. Beetge, R.W. Fearick, G. Mabala, S. Murray, N.J. Ncapayi, High-spin studies with the AFRODITE array, 1998.
[35] M. Lipoglavšek, A. Likar, M. Vencelj, T. Vidmar, R.A. Bark, E. Gueorguieva, F. Komati, J.J. Lawrie, S.M. Maliage, S.M. Mullins, et al., Measuring high-energy γ-rays with Ge clover detectors, Nucl. Instrum. Methods Phys. Res., Sect. A, Accel. Spectrom. Detect. Assoc. Equip. 557(2) (2006) 523–527.
[36] Y. Rahma, S. Ouichaoui, J. Kiener, E. A. Lawrie et al., Nuclear Physics A 1032 (2023) 122622, www.elsevier.com/locate/nuclphysa

[37] S. Ouichaoui, Y. Rahma, J. Kiener, E. A. Lawrie et al., 28th International Nuclear Physics Conference (INPC 2022), Journal of Physics: Conference Series 2586 (2023) 012099, IOP Publishing, https://doi:10.1088/1742-6596/2586/1/012099
[38] Y. Rahma, S. Ouichaoui, J. Kiener, E. A. Lawrie et al., 28th International Nuclear Physics Conference (INPC 2022), Journal of Physics: Conference Series 2586 (2023) 012099, IOP Publishing, https://doi:doi:10.1088/1742-6596/2586/1/012098
[39] W. Yahia Cherif, S. Damache, S. Ouichaoui, J. Kiener, E. A. Lawrie et al., Optical model potential parameter optimization for nucleon induced reactions on $^{40}$Ca: Implications on $\gamma$-ray production cross sections for residual Argon nuclei, EPJ Web of Conferences 322, 05004 (2025) CNR*24, https://doi.org/10.1051/epjconf/202532205004
[40] G. Duchêne, F.A. Beck, P.J. Twin, G. De France, D. Curien, L. Han, C.W. Beausang, M.A. Bentley, P.J. Nolan, J. Simpson, The clover: a new generation of composite Ge detectors, Nucl. Instrum. Methods Phys. Res., Sect. A, Accel. Spectrom. Detect. Assoc. Equip. 432(1) (1999) 90–110.
[41] M. Lipoglavšek, A. Likar, M. Vencelj, T. Vidmar, R.A. Bark, E. Gueorguieva, F. Komati, J.J. Lawrie, S.M. Maliage, S.M. Mullins, et al., Measuring high-energy γ-rays with Ge clover detectors, Nucl. Instrum. Methods Phys. Res., Sect. A, Accel. Spectrom. Detect. Assoc. Equip. 557(2) (2006) 523–527.
[42] J. Cresswell, J. Sampson, 2010•nnsa.dl.ac.uk, Mtsort language-edoc 033, 2010.
[43] ROOT-CERNLIB, https://root.cern.ch/
[44] R. C. Radford, https://radware.phy.ornl.gov/
[45] S. Agostinelli, J. Allison, K. Amako, J. Apostolakis, H. Araujo, P. Arce, M. Asai, D. Axen, S. Banerjee, G. Barrand, et al., GEANT4-a simulation toolkit, Nucl. Instrum. Methods Phys. Res., Sect. A, Accel. Spectrom. Detect. Assoc. Equip. 506(3) (2003) 250–303.
[46] M.E. Rose, The analysis of angular correlation and angular distribution data, Phys. Rev. 91(3) (1953) 610.
[47] C. Iliadis, Nuclear Physics of Stars, John Wiley & Sons, 2007.
[48] G. Gilmore, Practical Gamma-Ray Spectroscopy, 2nd edition, John Wiley & Sons, New York, 2008.
[49] R.L. Bunting, J.J. Kraushaar, Short-lived radioactivity induced in Ge (Li) gamma-ray detectors by neutrons, Nucl. Instrum. Methods 118(2) (1974) 565–572.
[50] J. Kiener, Shape and angular distribution of the 4.439-MeV γ-ray line from proton inelastic scattering off 12C, Phys. Rev. C 99(1) (2019) 014605.
[51] R. Capote, M. H erman, P. Obložinský, P. G. Young, S. Goriely, T. Belgya, A. V. Ignatyuk,

A. J. Koning, S. Hilaire, V. A. Plujko *et al.*, Nuclear Data Sheets 110, 3107 (2009).

[52] P. A. Moldauer, Physical Review C 14, 764 (1976).

[53] J. Raynal, CEA SaclayreportCEA-N-2772 (1994).

[54] J. Raynal, CEA SaclayreportCEA-N-2772 (1994).

[55] V.V. Zerkin and B. Pritychenko. Nucl. Instrum.
Methods Phys. Res. A 888, 31 (2018), https://doi.org/10.1016/j.nima.2018.01.045

[ 56] E.S. Soukhovitski et al., OPTMAN code v1. https://www-nds.iaea.org/RIPL-3/codes/OPTMAN/

[ 57] RIPL-3 library, https://www-nds.iaea.org/RIPL-3/

[ 58] M. Herman, R. Capote, B. Carlson, P. Obložinský, M. Sin, A. Trkov, H. Wienke, and V. Zerkin, Nucl. Data Sheets 108, 2655 (2007).

[59] Vincent Tatischeff, Jurgen Kiener, γ-ray lines from cosmic-ray interactions with interstellar dust grains, New Astronomy Reviews 48 (2004) 99–103.

[60] Siegert, T., Calore, F., Jean, P. *et al.* γ-Ray Lines – Signatures of Nucleosynthesis, Cosmic Rays, Positron Annihilation, and Fundamental Physics. *Space Sci Rev* 222, 34 (2026). https://doi.org/10.1007/s11214-026-01281-y

[61] https://science.nasa.gov/ems/12_gammarays/

[62]https://www.esa.int/Science_Exploration/Space_Science/Integral/Why_do_we_observe_gamma_rays

[63] Thomas Rauscher, Friedrich-Karl Thielemann, Karl-Ludwig Kratz, Nuclear level density and the determination of thermonuclear rates for astrophysics, Phys. Rev. C 56, 1613, Published 1 September, 1997 ; https://doi.org/10.1103/PhysRevC.56.1613

[64] https://www.mpa-garching.mpg.de/27937/stellar-astrophysics-old

[65] De Angelis, A., Tatischeff, V., Argan, A. *et al.* Gamma-ray astrophysics in the MeV range. *Exp Astron* 51, 1225–1254 (2021), https://doi.org/10.1007/s10686-021-09706-y

[66] https://www.mpg.de/8383537/supernova_gamma_ray_lines

[67] https://space-science.llnl.gov/research/gamma-ray-spectrometers

[68] V. A. Dogiel, V. Tatischeff, K. S. Cheng, D. O. Chernyshov1, C. M. Ko, Nuclear interaction gamma-ray lines from the Galactic center region, A&A 508, 1-7 (2009)

[69] Lu-Yao Jiang, Yun Wang, Yu-Jia Wei, Da-Ming Wei, Xiang Li, Hao-Ning He, Jia Ren, Zhao-Qiang Shen, Zhi-Ping Jin, Probable evidence for a transient mega-electron volt emission line in the GRB 221023A, Communications. 2025 Mar 18;16:2668, doi: 10.1038/s41467-025-57791-w

[70] Andrea Francesco Battaglia1 and Säm Krucker, New insights into the proton precipitation sites in solar flares, A&A, 694, A58 (2025).

[71] E. Tirado-Bueno 1, J. E. Mendoza-Torres 2 and C. Amador-Meléndez, High-Energy Neutron Enhancements Not Classified as GLEs: A Case Study Using Ground-Based Neutron Monitor Data, Revista Mexicana de Astronomia y Astrofisica, 61, 233-239, (2025).

[72] A. Casanovas-Hoste, C. Domingo-Pardo, J. Lerendegui-Marco, C. Guerrero, A. Tarifeño-Saldivia, M. Krtička, M. Pigna, tari, F. Calviño, D. Schumann et al., the nTOF Collaboration

[73] De Angelis, A., Tatischeff, V., Argan, A. *et al.*, Gamma-ray astrophysics in the MeV range, *Exp Astron* 51, 1225–1254 (2021), https://doi.org/10.1007/s10686-021-09706-y

[74] Bialy, S. Cold clouds as cosmic-ray detectors. *Commun Phys* 3, 32 (2020), https://doi.org/10.1038/s42005-020-0293-7

[75] Jack T. Wilson ,* David J. Lawrence , and Patrick N. Peplowski, Vincent R. Eke and Jacob A. Kegerreis, Space-based measurement of the neutron lifetime using data from the neutron spectrometer, PHYSICAL REVIEW RESEARCH 2, 023316 (2020), https://doi.org/10.1103/PhysRevResearch.2.023316

[76] Randall K. Smith, Nancy S. Brickhouse, Chapter Four - Atomic Data Needs for Understanding X-ray Astrophysical Plasmas, Advances In Atomic, Molecular, and Optical Physics Volume 63, 2014, Pages 271-321, https://doi.org/10.1016/B978-0-12-800129-5.00004-3

[77] Siegert, T., Calore, F., Jean, P. *et al.* -Ray Lines – Signatures of Nucleosynthesis, Cosmic Rays, Positron Annihilation, and Fundamental Physics. *Space Sci Rev* 222, 34 (2026), https://doi.org/10.1007/s11214-026-01281-y

[78] Thomas Siegert, Sohan Ghosh, Kalp Mathur, Ethan Spraggon and Akshay Yeddanapudi, Nucleosynthesis constraints through $\gamma$-ray line measurements from classical novae. Hierarchical model for the ejecta of $^{22}Na$ and $^{7}Be$, A&A, Vol. 650 (2021) A187, https://doi.org/10.1051/0004-6361/202140300

[79] Amina Trabelsi, Mourad Fouka, Saad Ouichaoui and Amel Belhout, Modeling of the Gamma Ray Burst photospheric emission : Monte Carlo simulation of the GRB prompt emission, numerical results and discussionn, Astrophysics and Space Science (2024) 369:102, https://doi.org/10.1007/s10509-024-04366-8

[80] Yassine Rahmani, Abdelaziz Sid, Mourad Fouka, Saad Ouichaoui and RedouaneMecheri, Multibands fitting of Gamma-ray burst's afterglow's light curves using the synchrotron external forward shock model, Astrophysics and Space Science (2024) 369:15, https://doi.org/10.1007/s10509-024-04279-6

[81] N. Indriolo, B. D. Fields and B. J. McCall, Apj. **694** (2009) 257.

[82] Y. Shikaze et *al*, Astroparticle Phys. **28** (2007) 154.

[83] A. Bhatnagar, W. Livingstone, ''*Fundamentals of solar astronomy*'', World Scientific Publishing (2005).